\documentclass[epj]{svjour}
\usepackage{graphics,graphicx}
\usepackage{rotating}
\begin{document}
\hugehead
\title{Exclusive $\rho^0$ electroproduction on the proton at CLAS}

\newcommand*{\ANL}{Argonne National Laboratory, Illinois 60439}
\institute{\ANL}
\newcommand*{\ASU}{Arizona State University, Tempe, Arizona 85287-1504}
\institute{\ASU}
\newcommand*{\UCLA}{University of California at Los Angeles, Los Angeles, California  90095-1547}
\institute{\UCLA}
\newcommand*{\CSU}{California State University, Dominguez Hills, Carson, CA 90747}
\institute{\CSU}
\newcommand*{\CMU}{Carnegie Mellon University, Pittsburgh, Pennsylvania 15213}
\institute{\CMU}
\newcommand*{\CUA}{Catholic University of America, Washington, D.C. 20064}
\institute{\CUA}
\newcommand*{\SACLAY}{CEA-Saclay, Service de Physique Nucl\'eaire, 91191 Gif-sur-Yvette, France}
\institute{\SACLAY}
\newcommand*{\CNU}{Christopher Newport University, Newport News, Virginia 23606}
\institute{\CNU}
\newcommand*{\UCONN}{University of Connecticut, Storrs, Connecticut 06269}
\institute{\UCONN}
\newcommand*{\ECOSSEE}{Edinburgh University, Edinburgh EH9 3JZ, United Kingdom}
\institute{\ECOSSEE}
\newcommand*{\EMMY}{Emmy-Noether Foundation, Germany}
\institute{\EMMY}
\newcommand*{\FU}{Fairfield University, Fairfield CT 06824}
\institute{\FU}
\newcommand*{\FIU}{Florida International University, Miami, Florida 33199}
\institute{\FIU}
\newcommand*{\FSU}{Florida State University, Tallahassee, Florida 32306}
\institute{\FSU}
\newcommand*{\GWU}{The George Washington University, Washington, DC 20052}
\institute{\GWU}
\newcommand*{\ECOSSEG}{University of Glasgow, Glasgow G12 8QQ, United Kingdom}
\institute{\ECOSSEG}
\newcommand*{\ISU}{Idaho State University, Pocatello, Idaho 83209}
\institute{\ISU}
\newcommand*{\INFNFR}{INFN, Laboratori Nazionali di Frascati, 00044 Frascati, Italy}
\institute{\INFNFR}
\newcommand*{\INFNGE}{INFN, Sezione di Genova, 16146 Genova, Italy}
\institute{\INFNGE}
\newcommand*{\ORSAY}{Institut de Physique Nucleaire ORSAY, Orsay, France}
\institute{\ORSAY}
\newcommand*{\BONN}{Institute f\"{u}r Strahlen und Kernphysik, Universit\"{a}t Bonn, Germany}
\institute{\BONN}
\newcommand*{\ITEP}{Institute of Theoretical and Experimental Physics, Moscow, 117259, Russia}
\institute{\ITEP}
\newcommand*{\JMU}{James Madison University, Harrisonburg, Virginia 22807}
\institute{\JMU}
\newcommand*{\KYUNGPOOK}{Kyungpook National University, Daegu 702-701, Republic of Korea}
\institute{\KYUNGPOOK}
\newcommand*{\MIT}{Massachusetts Institute of Technology, Cambridge, Massachusetts  02139-4307}
\institute{\MIT}
\newcommand*{\UMASS}{University of Massachusetts, Amherst, Massachusetts  01003}
\institute{\UMASS}
\newcommand*{\MOSCOW}{Moscow State University, General Nuclear Physics Institute, 119899 Moscow, Russia}
\institute{\MOSCOW}
\newcommand*{\UNH}{University of New Hampshire, Durham, New Hampshire 03824-3568}
\institute{\UNH}
\newcommand*{\NSU}{Norfolk State University, Norfolk, Virginia 23504}
\institute{\NSU}
\newcommand*{\OHIOU}{Ohio University, Athens, Ohio  45701}
\institute{\OHIOU}
\newcommand*{\ODU}{Old Dominion University, Norfolk, Virginia 23529}
\institute{\ODU}
\newcommand*{\PITT}{University of Pittsburgh, Pittsburgh, Pennsylvania 15260}
\institute{\PITT}
\newcommand*{\RPI}{Rensselaer Polytechnic Institute, Troy, New York 12180-3590}
\institute{\RPI}
\newcommand*{\RICE}{Rice University, Houston, Texas 77005-1892}
\institute{\RICE}
\newcommand*{\URICH}{University of Richmond, Richmond, Virginia 23173}
\institute{\URICH}
\newcommand*{\SCAROLINA}{University of South Carolina, Columbia, South Carolina 29208}
\institute{\SCAROLINA}
\newcommand*{\JLAB}{Thomas Jefferson National Accelerator Facility, Newport News, Virginia 23606}
\institute{\JLAB}
\newcommand*{\UNIONC}{Union College, Schenectady, NY 12308}
\institute{\UNIONC}
\newcommand*{\NONE}{unknown}
\institute{\NONE}
\newcommand*{\VT}{Virginia Polytechnic Institute and State University, Blacksburg, Virginia   24061-0435}
\institute{\VT}
\newcommand*{\VIRGINIA}{University of Virginia, Charlottesville, Virginia 22901}
\institute{\VIRGINIA}
\newcommand*{\WM}{College of William and Mary, Williamsburg, Virginia 23187-8795}
\institute{\WM}
\newcommand*{\YEREVAN}{Yerevan Physics Institute, 375036 Yerevan, Armenia}
\institute{\YEREVAN}
\newcommand*{\VAL}{Universidad T\'ecnica Federico Santa Mar\'{\i}a,
Casilla 110-V, Valpara\'\i so, Chile}
\institute{\VAL}
\newcommand*{\NOWOHIOU}{Ohio University, Athens, Ohio  45701}
\newcommand*{\NOWJLAB}{Thomas Jefferson National Accelerator Facility, Newport News, Virginia 23606}
\newcommand*{\NOWUNH}{University of New Hampshire, Durham, New Hampshire 03824-3568}
\newcommand*{\NOWCNU}{Christopher Newport University, Newport News, Virginia 23606}
\newcommand*{\NOWGWU}{The George Washington University, Washington, DC 20052}
\newcommand*{\NOWUMASS}{University of Massachusetts, Amherst, Massachusetts  01003}
\newcommand*{\NOWMIT}{Massachusetts Institute of Technology, Cambridge, Massachusetts  02139-4307}
\newcommand*{\NOWURICH}{University of Richmond, Richmond, Virginia 23173}
\newcommand*{\NOWECOSSEE}{Edinburgh University, Edinburgh EH9 3JZ, United Kingdom}

\author{S.A.~Morrow~\inst{1,2}
\and M.~Guidal~\inst{1}\footnote{\textit{Corresponding author:} guidal@ipno.in2p3.fr} 
\and M.~Gar\c con~\inst{2}
\and J.M.~Laget~\inst{2,3}
\and E.S.~Smith~\inst{3}
\and G.~Adams~\inst{4} 
\and K.P.~Adhikari~\inst{5}
\and M.~Aghasyan~\inst{6}
\and M.J.~Amaryan~\inst{5}
\and M.~Anghinolfi~\inst{7}
% ...
	\\
	\centerline{(CLAS Collaboration)}
}
\clearpage

\institute{\ORSAY
\and\SACLAY
\and\JLAB
\and\RPI
\and\ODU
\and\INFNFR
\and\INFNGE
% ...
}

\date{Received: date / Revised version: date}
%
\abstract{The $e p\to e^\prime p \rho^0$ reaction has been measured,
using the 5.754 GeV electron beam of Jefferson Lab and the CLAS detector.
This represents the largest ever set of data for this reaction in the valence 
region. Integrated and differential cross sections are presented.
The $W$, $Q^2$ and $t$ dependences of the cross section are compared 
to theoretical calculations based on $t$-channel meson-exchange Regge 
theory on the one hand and on quark handbag diagrams related to Generalized
Parton Distributions (GPDs) on the other hand. The Regge 
approach can describe at the $\approx$ 30\% level most of the features 
of the present data while the two GPD calculations that are
presented in this article which succesfully reproduce the high energy data 
strongly underestimate the present data. The question is then raised whether 
this discrepancy originates from an incomplete or inexact way of modelling the 
GPDs or the associated hard scattering amplitude or whether the GPD formalism is 
simply inapplicable in this region due to higher-twists contributions,
incalculable at present.
} 
\maketitle
\section{Introduction}
\label{intro}

The exclusive electroproduction of photons and mesons on the nucleon
is an important tool to better understand nucleon structure and, 
more generally, the transition between the low energy hadronic 
and high energy partonic domains of the Quantum Chromodynamics (QCD) theory.

\begin{figure}[h]
\includegraphics[width=1.\columnwidth,bb=75 375 550 600,clip=true]{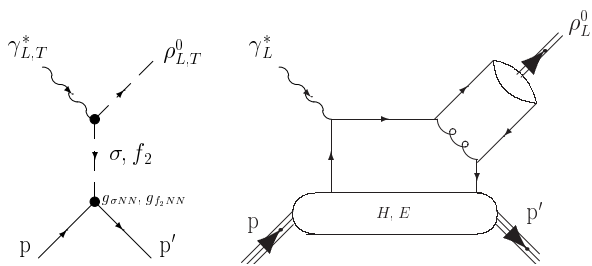}
\caption{The mechanisms for $\rho^0$ electroproduction at JLab
energies for low $Q^2$ (left diagram) through the exchange of mesons and
for high $Q^2$ (right diagram) through the quark exchange ``handbag" mechanism
(valid for longitudinal photons) where $H$ and $E$ are the unpolarized GPDs.}
\label{diags}
\end{figure}

Among all such exclusive processes, the $e p\to e^\prime p \rho^0$ reaction
bears some particular advantages. It is a process for which numerical
calculations and predictions are available both in terms of $hadronic$ degrees of 
freedom, via Reggeized meson exchanges, and in terms of $partonic$ degrees of freedom,
via Generalized Parton Distributions (GPDs). We refer the reader to
refs.~\cite{Regge,collinsreg,storrow} and refs.~\cite{muller,ji,rady,collins,goeke,revdiehl,revrady} 
for the original articles and general reviews of Regge theory and GPDs respectively.
Defining $Q^2$ as the absolute value of the squared mass of the virtual photon
that is exchanged between
the electron and the target nucleon, partonic descriptions are expected to be valid at large  
$Q^2$, while hadronic descriptions dominate in photo- and low-$Q^2$ 
electroproduction. Fig.~\ref{diags} illustrates these two approaches at the electron beam energies
available at Jefferson Laboratory (JLab).
Concerning the Reggeized meson exchange approach, the total and differential cross sections 
associated with
the exchanges of the dominant Regge $\sigma$ and $f_2$ trajectories have been calculated 
by Laget {\it et al.}~\cite{RgModel2,RgModel3}. Concerning the GPDs approach, the so-called ``handbag" diagram,
with recent modelings of the
unpolarized GPDs, has been calculated by two groups: Goloskokov-Kroll~\cite{gk}
and Vanderhaeghen {\it et al.}~\cite{Vdh1,Vdh2,Vdh3,Vdh4}.
Let us note here that in the GPD approach the leading twist handbag calculation is valid only 
for the longitudinal part of the cross section and that, experimentally,
it is important to separate the longitudinal and transverse parts of the 
cross sections when measuring this process.

This article presents results for the exclusive electroproduction 
of the $\rho^0$ vector meson on the proton measured with the 5.754 GeV 
electron beam of the CEBAF accelerator and the CEBAF Large Acceptance
Spectrometer (CLAS) at JLab. 
The aim of this analysis is to compare the integrated and differential 
cross sections of the $e p\to e^\prime p \rho^0$ 
reaction that have been extracted over the intermediate $Q^2$ region accessible 
at CLAS, with the two Regge and GPD
theoretical approaches, and thus determine their domain of validity and constrain 
their various inputs.

There are a few existing electroproduction data in a similar kinematical regime:
early data with the 7.2 GeV beam at DESY~\cite{JoosRho}
and with the 11.5 GeV beam at Cornell~\cite{Cassel}, and more
recently with the 27 GeV beam at HERMES~\cite{HERMESrho} and
the 4.2 GeV beam of JLab~\cite{cynthia}. 
The present work explores new phase-space regions and,
in regions of overlap, has much finer binning and precision.

In section~\ref{exppro} we present the experimental procedure we
have adopted to extract our integrated and differential cross sections. 
In section~\ref{theo}, after briefly describing the 
Regge and GPDs models, we compare these calculations to our data.
Finally, we draw our conclusions in section~\ref{conc}.

\section{Experimental procedure}
\label{exppro}

The CLAS detector~\cite{mecking} is built around six superconduction coils
that generate a toroidal magnetic field primarily in the azimuthal
direction. Each sector is equipped with three regions of multi-wire drift
chambers (DC) and time-of-flight scintillator counters (SC) that cover 
the angular range from 8$^\circ$ to 143$^\circ$. In the forward region 
(8$^\circ$$<$ $\theta$ $<$ 45$^\circ$), each sector is furthermore equipped
with a gas-filled threshold Cerenkov counter (CC) and a lead-scintillator 
sandwich type electromagnetic calorimeter (EC). Azimuthal coverage for 
CLAS is limited by the magnet's six coils and is approximatively 90\% 
at large polar angles and 50\% at forward angles. 

The data were taken with an electron beam having an energy of 5.754 GeV impinging on 
an unpolarized 5-cm-long liquid-hydrogen target.
The integrated luminosity of this data set was 28.5 fb$^{-1}$.
The data were taken from October 2001 to January 2002.
The kinematic domain of the selected sample corresponds approximately to
$Q^2$ from 1.5 to 5.5 GeV$^2$. We analyzed data with
$W$, the $\gamma^*-p$ center-of-mass energy, greater than 1.8 GeV, which 
corresponds to a range of $x_B$ approximatively from 0.15 to 0.7.
Here $x_B$ is the standard Bjorken variable equal to $\frac{Q^2}{W^2-m^2_p+Q^2}$ 
with $m_p$ the mass of the proton.

The $\rho^0$ decays into two pions ($\pi^+\pi^-$), with a branching
ratio of~100\%~\cite{PDG}. To select the channel $ep \rightarrow
e^\prime p\rho^0$, we based our analysis on the identification of 
the scattered electron, the recoil proton and the positive decay pion
(because of the polarity of the magnetic field, negative pions
are bent toward the beam pipe and in general escape the acceptance
of CLAS); we then used the missing mass $ep\to e^\prime p\pi^+X$ for 
the identification of the 
$ep \rightarrow e^\prime p\pi^+\pi^-$ final state. 

Once this final state is identified and its yield normalized, the reduced 
$\gamma^*p \rightarrow p\rho^0$ cross section is extracted 
by fitting in a model-dependent way the ($\pi^+\pi^-$) invariant mass
using a parametrized $\rho^0$ shape, which will be described later. 
The longitudinal and transverse
cross sections are then further extracted by analyzing the decay pion 
angular distribution in the $\rho^0$ center-of-mass frame. We detail 
all these steps in the following sections.

\subsection{Particle identification}
\label{partid}

The electron is identified as a negative track, determined from the DC, 
having produced a signal in the CC and the EC. Pions, potentially 
misidentified as electrons, were rejected by cutting on the CC amplitude
($>$ 2 photoelectrons), imposing a minimum energy deposition in 
the EC (60 MeV) and correlating the measurements of
the momentum from the DC and of the energy from the EC. 
In order to minimize radiative corrections and residual pion contamination,
a further cut E$^\prime\geq$~0.8~GeV was also applied, where E$^\prime$ is 
the scattered electron energy. Finally, vertex and geometric fiducial cuts,
which select only regions of well understood acceptance, were included. 

The efficiencies of the CC and EC cuts, respectively $\eta_{CC}$ and 
$\eta_{EC}$, were determined from data samples, selecting unambiguous electrons 
with very tight CC or EC cuts. The CC-cut efficiencies range from 86 to 99\% and 
the EC-cut efficiencies from 90 to 95\%, depending on the electron kinematics.
The efficiencies of the geometric fiducial cuts were derived from CLAS GEANT-based 
Monte-Carlo simulations.

Pions and protons are identified by the correlation between the 
momentum measured by the DC and the velocity measured by the SC.
This identification procedure is unambiguous for particles
with momenta up to 2 GeV. Particles with momenta higher than 2 GeV
were therefore discarded. The efficiencies of the cuts imposed for
this momentum-velocity correlation and of the geometric fiducial cuts were
determined from CLAS GEANT-based Monte-Carlo simulations.

\begin{figure}[h]
\includegraphics[width=8cm]{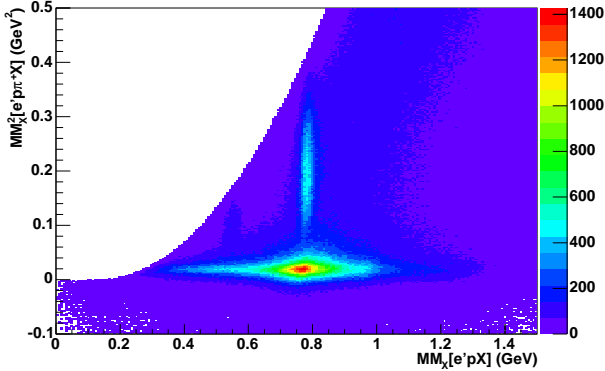}
\caption{Squared missing mass $M_{X}^2 [e^\prime p\pi^+X]$ vs $M_{X} [e^\prime pX]$ for $W \geq$~1.8~GeV 
         and $E^\prime\geq$~0.8~GeV.}
\label{fig:mm2dim}
\end{figure}

\begin{figure}[h!]
\begin{center}
\includegraphics[width=8cm]{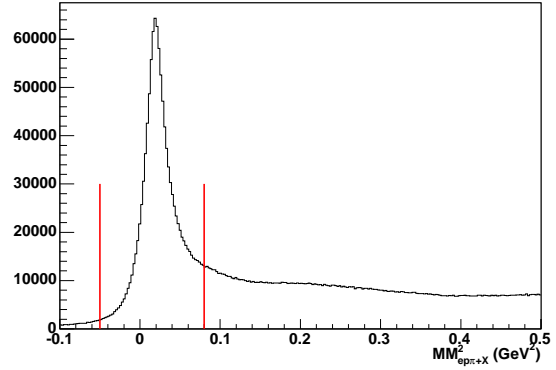}
\caption{Missing mass $M_{X}^2 [e^\prime p\pi^+X]$ for $W \geq$~1.8~GeV 
         and $E^\prime\geq$~0.8~GeV. The red lines show the cut used 
	 (see eq.~\ref{eq:mm2eppipcut}) to select the $e^\prime p\pi^+\pi^-$final state.}
\label{fig:mm2eppip}
\end{center}
\end{figure}

Once the electron, the proton and the positive pion are identified, the 
$ep \rightarrow e^\prime p\pi^+\pi^-$ final state is identified through the 
missing mass technique.
Fig.~\ref{fig:mm2dim} shows the square of the missing mass for the system 
$e^\prime p\pi^+$ (i.e. $M_{X}^2 [e^\prime p\pi^+X]$) as a function of the 
missing mass for the system $ep$ (i.e. $M_{X} [e^\prime pX]$). One distinguishes 
the $\rho^0$ and $\omega$ loci quite clearly. A cut on the 
$M_{X}^2 [e^\prime p\pi^+X]$ variable is required in order to separate 
the $\rho^0$ and the associated $\pi^+\pi^-$ continuum from the 
$\omega$ and the three--pion continuum background.
The optimum value of this cut:
\begin{equation}
-0.05 \le M_{X}^2 [e^\prime p\pi^+X] \leq 0.08 \ \hbox{GeV}^2
\label{eq:mm2eppipcut}
\end{equation}
was determined from a study whereby we estimated the number of 
$\rho^0$ events, from fits to the $M_{X}^2 [e^\prime pX]$ distribution, 
as a function of the cut values. The cuts were chosen in the region where 
the number of $\rho^0$ events began to vary 
only very weakly with these cut values.The 
simulation used to calculate acceptances 
reproduces the features of fig.~\ref{fig:mm2eppip}.
The position of this cut relative to $M_{X}^2 [e^\prime p\pi^+X]$ is shown
in fig.~\ref{fig:mm2eppip}.

\begin{figure}[h!]
\begin{center}
\includegraphics[width=10cm]{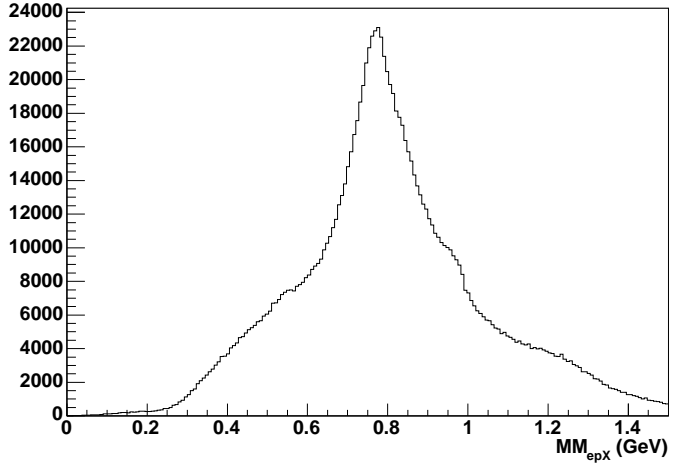}
\caption{Missing mass $M_{X} [e^\prime pX]$ for -0.05~$\leq M_{X}^2 [e^\prime p\pi^+] \leq$~0.08~GeV$^2$, 
	 $W \geq$~1.8~GeV
         and $E^\prime\geq$~0.8~GeV. The $\rho^0(770)$, as well as the
	 $f_0(980)$ and $f_2(1270)$ resonances which can be distinguished,
	 sits on top of a background of non-resonant two-pion continuum.}
\label{fig:mmep}
\end{center}
\end{figure}

The missing mass distribution for the system $ep$, obtained after this cut,
is shown in fig.~\ref{fig:mmep}. 
The $\rho^0$ peak is very 
broad~: $\Gamma_{\rho^0}^{th} \approx$ 150~MeV~from ref.~\cite{PDG}. 
and sits on top of a background of a non-resonant two pion continuum,
which originates from other processes leading to the $e^\prime p\pi^+\pi^-$  
final state, such as 
$ep \rightarrow e^\prime \Delta^{++}\pi^- \rightarrow e^\prime p\pi^+\pi^-$.
In fig.~\ref{fig:mmep}, one can additionnally distinguish two bumps 
at masses around 950 MeV and 1250 MeV corresponding respectively to 
the scalar $f_0(980)$ and tensor $f_2(1270)$ mesons. These will be even more
evident when we look at the differential spectra later on.

\subsection{Acceptance calculation}
\label{accep}

The acceptance of the CLAS detector for the $e^\prime p\pi^+X$ process has
been determined with the standard GEANT-based code developed for CLAS.
Our event generator~\cite{genev} contains the three main channels 
leading to the $e^\prime p\pi^+\pi^-$ final state: 
$ep \rightarrow e^\prime p\rho^0 \hookrightarrow \pi^+\pi^-$, 
$ep \rightarrow e^\prime \pi^-\Delta^{++} \hookrightarrow p\pi^+$, 
and the non-resonant (phase space) $ep \rightarrow e^\prime p\pi^+\pi^-$. 
This event generator is based on tables of total and differential cross 
sections of double pion photoproduction data that have been 
extrapolated to electroproduction. This has been done by multiplying these
tables by a virtual photon flux factor and a dipole form factor
in order to obtain a relatively realistic $Q^2$ dependence of
the cross section. We also have tuned the relative weight of all the
aforementioned channels in order to reproduce the main kinematical
distributions of our experimental data.

\begin{figure}[h!]
\begin{center}
\includegraphics[width=8cm]{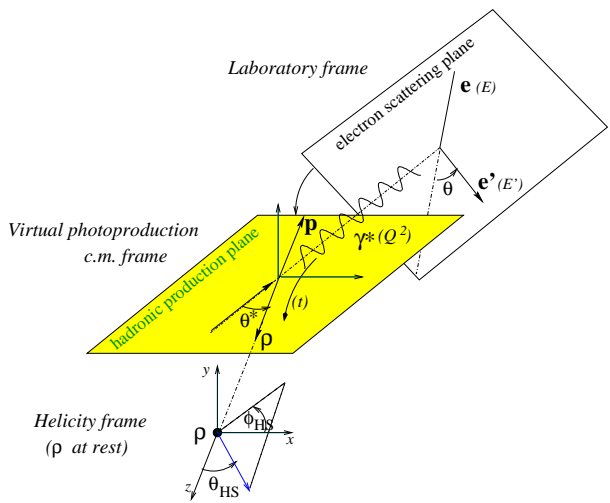}
\caption{Reference frames and relevant variables for the description
of the $ep \rightarrow e^\prime p\rho^0 \hookrightarrow \pi^+\pi^-$ reaction.}
\label{fig:rho0_kine}
\end{center}
\end{figure}

Eight independent kinematical variables are necessary to describe a reaction
with four particles in the final state. However, in unpolarized electroproduction,
the cross section does not depend on the azimuthal angle of the scattered
electron. The following seven variables are then chosen: 
$Q^2$, $x_B$, $t$, $M_{\pi^+\pi^-}$, $\Phi$, $\cos(\theta_{HS})$
and $\phi_{HS}$. Here $Q^2$ and $x_B$ are
respectively the absolute value of the squared electron four-momentum transfer and 
the Bjorken variable, which describe the kinematics of the virtual photon 
$\gamma^*$. At some stages, $W$, the $\gamma^*-p$ center-of-mass energy will
equivalently be used. Then $t$ is the squared four-momentum transferred to the $\rho^0$,
$\Phi$ is the azimuthal angle between the electron scattering plane and 
the hadronic production plane,
and $M_{\pi^+\pi^-}$ is the invariant mass of the $\pi^+\pi^-$ system.
Finally, $\cos(\theta_{HS})$ and $\phi_{HS}$ describe the decay of the
$\rho^0$ into two pions and are respectively the polar and azimuthal 
angles of the $\pi^+$ in the so-called Helicity Frame (HS) where the 
$\rho^0$ is at rest and the $z$-axis is given by the $\rho^0$ direction 
in the $\gamma^*-p$ center-of-mass system.
All these variables are illustrated in fig.~\ref{fig:rho0_kine}.

\begin{figure}[h!]
\begin{center}
\includegraphics[width=9cm]{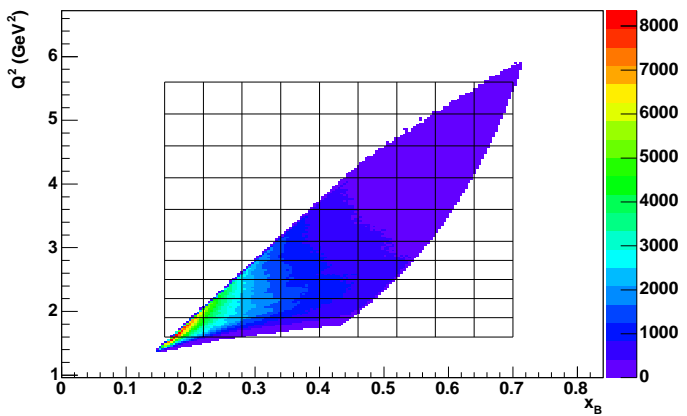}
\caption{Binning in ($Q^2$,$x_{B}$) for our experimental data 
(with $W>1.8$~GeV and $E^\prime\geq$~0.8~GeV).}
\label{fig:binning}
\end{center}
\end{figure}

The procedure we have followed has been to calculate an acceptance for each of the 
7-dimensional bins. In the limit of small bin-size
and unlimited statistics, this procedure is independent of the 
model used to generate events. Our event generator 
has nonetheless been tuned to reproduce the experimental data.
The binning in the 7 independent variables is defined in 
table~\ref{tab:binning} and its 2-dimensional ($Q^2$, $x_B$) projection
is shown in fig.~\ref{fig:binning}.

\begin{table}[h!]	
\begin{center}	

\begin{tabular}{|l|l|c|c|r|}
\hline	
Variable & Unit  & Range            & \# of bins & Width \\
\hline	
$Q^2$ & GeV$^2$  & 1.60 --   3.10   & 5 &  0.30 \\
      &          & 3.10 --   5.60   & 5 &  0.50 \\
\hline	
$x_{B}$ & --      & 0.16 --   0.7    & 9 &  0.06 \\
\hline		
$-t$ & GeV$^2$   & 0.10 --   1.90   & 6 &  0.30 \\
     &           & 1.90 --   4.30   & 3 &  0.80 \\
\hline	
$\Phi$ & deg. & 0.00 -- 360.00   & 9 & 40.00 \\
\hline
$\cos(\theta^{HS}_{\pi+})$ & --            & -1.00 --   1.00  &  8 &  0.25 \\
\hline
$\phi^{HS}_{\pi+}$         & deg.       &  0.00 -- 360.00  &  8 & 45.00 \\
\hline
MM$_{\rm ep}$              & GeV           &  0.22 --   1.87  & 15 &  0.22 \\
\hline	
\end{tabular}	
\caption{Binning in the 7 independent variables for the acceptance table.} 
\label{tab:binning}
\end{center}	
\end{table}	

More than 200 million Monte-Carlo (MC) events were generated using CLAS GEANT
to calculate the acceptance in each of the individual 7-dimensional bins.
Each ``real data" event was then weighted by the ratio 
of the number of MC generated to accepted events for each 7-dimensional bin. 
The acceptances are at most a few percent. Bins that have 
a very small acceptance ($<$ 0.16\%) have a very high 
weight, which produces an unphysically high and narrow peak in the weighted 
event distributions, and have to be cut away. The efficiency 
of this cut is evaluated by MC computation of the ratio of weighted accepted
events to generated events mapped onto 1-dimensional distributions. This correction 
factor $\eta_{w}$ is therefore model dependent since it is 1-dimensional and thus 
integrated over the remaining variables. It is on average 15\%.

Radiative corrections were part of our event generator and 
were calculated according to ref.~\cite{RadCorr}. The MC acceptance
calculation presented above therefore took into account the effects
of the radiation of hard photons and the corresponding losses due
to the application of the cut of eq.~\ref{eq:mm2eppipcut}.
The contribution of soft photons and virtual corrections were determined by turning
on and off the radiative effects in our event generator, defining
an $F_{rad}$ factor for each ($Q^2$, $x_B$) bin for the integrated cross sections, 
or for each ($Q^2$, $x_B$, $X$) bin for the differential cross sections, 
$X$ being one of $t$, $\Phi$, $\cos(\theta_{HS})$ or $\phi_{HS}$.

\subsection{$\gamma^*p \rightarrow p\pi^+\pi^-$ total cross section}

The total reduced cross section for the $ep \rightarrow e^\prime p\pi^+\pi^-$ 
reaction can then be obtained from:

\begin{equation}
\sigma_{\gamma^* p \rightarrow p\pi^+\pi^-} (Q^2, x_B, E) = 
\frac{1}{\Gamma_V(Q^2,x_B,E)} \frac{d^2\sigma_{ep\rightarrow e^\prime p\pi^+\pi^-}}
{dQ^2 dx_B}
\end{equation}

\noindent with:
\begin{equation}
\frac{d^2 \sigma_{ep \rightarrow e^\prime p\pi^+\pi^-}}{dQ^2 dx_B} =
\frac{n_{w}(Q^2,x_B)}{\mathcal{L}_{int} \ \Delta Q^2 \ \Delta x_B} \times
\frac{F_{rad}}{\eta_{CC} \eta_{EC} \eta_{w}},
\label{eq:sig_ep_eppippim}
\end{equation}

\noindent where
\begin{itemize}
\item $n_{w} (Q^2,x_B)$ is the weighted number of 
	$ep \rightarrow e^\prime p\pi^+\pi^-$ events in a given bin ($Q^2$, $x_B$),
\item $\mathcal{L}_{int}$ is the effective integrated luminosity (that takes
into account the correction for the data acquisition dead time),
\item $\Delta Q^2$ and $\Delta x_B$ are the corresponding bin widths 
	(see table~\ref{tab:binning}); 
	for bins not completely filled, because of $W$ or $E^\prime$ cuts on 
	the electron for instance (see fig.~\ref{fig:binning}), 
	the phase space $\Delta Q^2 \Delta x_B$
	includes a surface correction and the $Q^2$ and $x_B$ central
	values are modified accordingly.
\item $F_{rad}$ is the correction factor due to the radiative effects
(see section~\ref{accep}),
\item $\eta_{CC}$ is the CC-cut efficiency (see section~\ref{partid}),
\item $\eta_{EC}$ is the EC-cut efficiency (see section~\ref{partid}),
\item $\eta_{w}$ is the efficiency of the cut on the weight in
the acceptance calculation (see section~\ref{accep}). \\
\end{itemize}

We adopted the Hand convention~\cite{Hand} for the
definition of the virtual photon flux $\Gamma_V$:

\begin{equation}
\Gamma_V (Q^2,x_B,E) = \frac{\alpha}{8\pi} \frac{Q^2}{m_p^2 E^2} 
\frac{1-x_B}{x_B^3} \frac{1}{1-\epsilon}
\label{eq:GammaV}
\end{equation}
with 
\begin{equation}
\epsilon = \frac{1}{1+2\frac{Q^2+(E-E^\prime)^2}{4EE^\prime-Q^2}}
\label{eq:epsilon}
\end{equation}

\noindent and $\alpha\approx\frac{1}{137}$ the standard electromagnetic coupling constant.

\begin{figure*}
\begin{center}
\includegraphics[width=15cm]{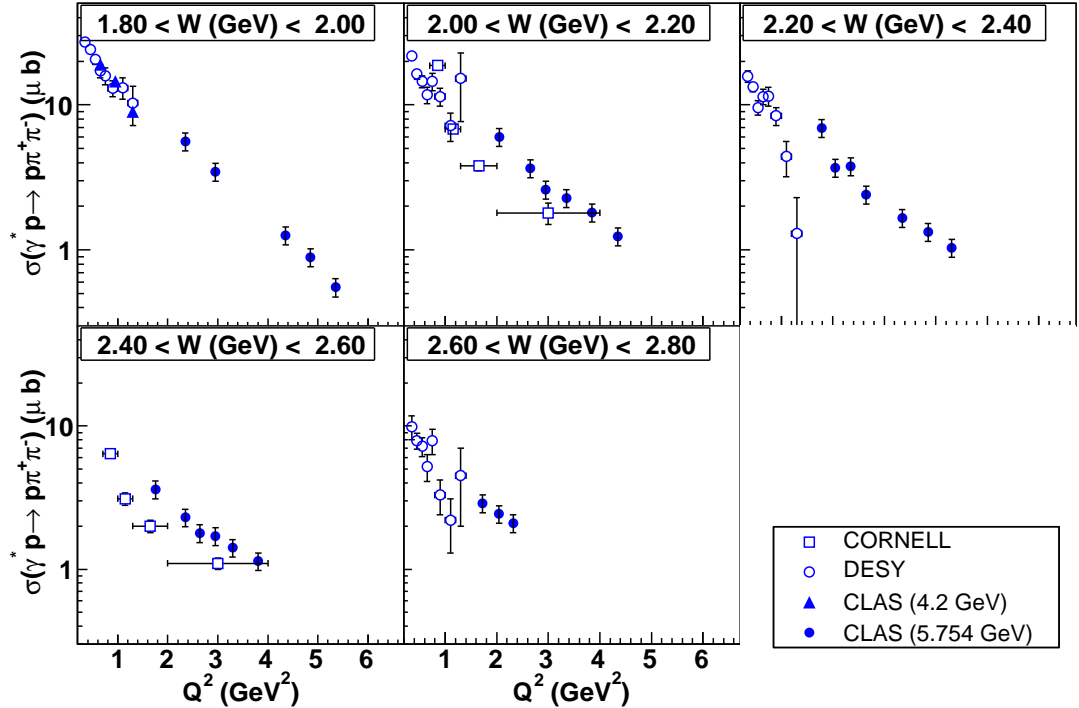}
\caption{Reduced cross sections $\gamma^* p \rightarrow p \pi^+ \pi^-$ 
	as a function of $Q^2$ for constant $W$ bins, 
	in units of $\mu$barn from the current analysis. Also shown are earlier 
	data from CLAS with a 4.2 GeV beam energy~\cite{ripani} as well as the
	data from DESY~\cite{JoosRho} and Cornell~\cite{Cassel} with, respectively,
	a 7.2 GeV and 11.5 GeV beam energy.}
\label{fig:XsectPipPimCompWorldvQ2}
\end{center}
\end{figure*}

Fig.~\ref{fig:XsectPipPimCompWorldvQ2} 
shows the total reduced cross section 
$\sigma_{\gamma^*p \rightarrow p \pi^+ \pi^-}$ as a function of 
$Q^2$ for constant $W$ bins 
compared with the world's data~\cite{JoosRho,Cassel,ripani}.

Relatively good agreement between the various experiments can be seen.
It is important to realize that what is plotted is the unseparated 
cross section, i.e. a linear combination of the transverse ($\sigma_T$) 
and longitudinal ($\sigma_L$) cross sections~: 
$\sigma=\sigma_T+\epsilon\sigma_L$. This means that, due to $\epsilon$ 
(eq.~\ref{eq:epsilon}), 
there is a dependence on the beam energy in this observable.
Since the CORNELL data have been taken with an 11.5 GeV electron 
beam energy~\cite{Cassel}, the DESY data with a 7.2 GeV electron 
beam energy~\cite{JoosRho} and the previous CLAS data with a 4.2 GeV beam~\cite{ripani}, 
the data sets, although at approximatively
equivalent $Q^2$ and $W$ values, are not directly comparable
and are not expected to fully match each other. We will come back to this
issue in section~\ref{int} when we are comparing the 
$\rho^0$ cross sections.

The next step is to extract the $\gamma^*p \rightarrow p \rho^0$ cross 
section from the $\gamma^*p \rightarrow p \pi^+ \pi^-$ cross 
section, which requires a dedicated fitting procedure.

\subsection{Fitting procedure for the $\gamma^*p \rightarrow p\rho^0$ 
cross section}
\label{back}

\begin{figure*}
\begin{center}
\includegraphics[width=18cm]{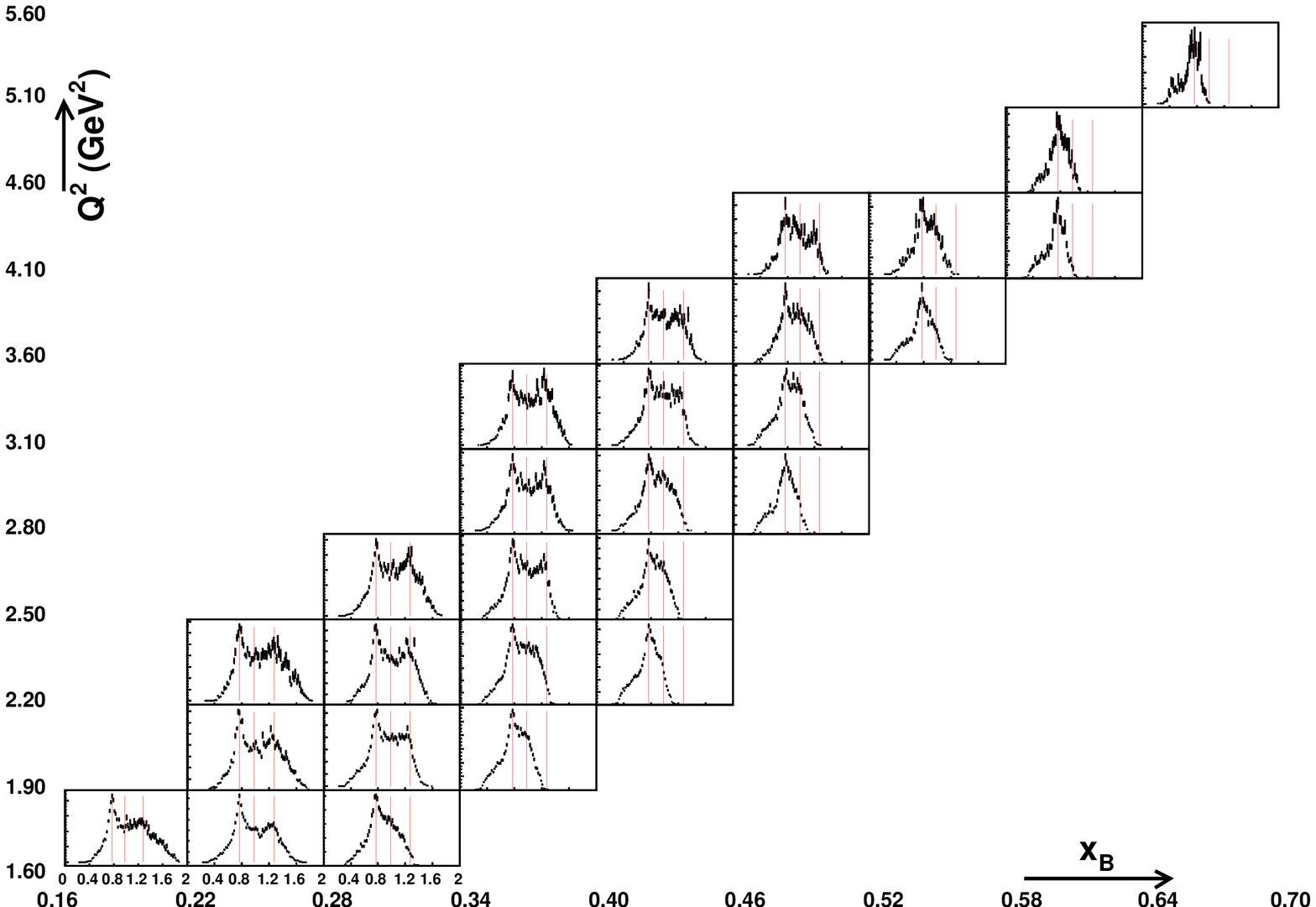}
\caption{Acceptance-corrected $M_{X} [e^\prime pX]$ (in GeV) missing mass spectra for all
our ($Q^2$, $x_B$) bins. The three red lines are located at $M_{X}$ = 0.770, 0.980, 1.275 GeV
corresponding to the three well-known resonant states in the ($\pi^+\pi^-$) system.}
\label{fig:MMepXWeighted}
\end{center}
\end{figure*}

Fig.~\ref{fig:MMepXWeighted} shows the acceptance-weighted $M_{\pi^+\pi^-}$ 
spectra for all our ($Q^2$,$x_B$) bins. The $\rho^0$ peak (along with the 
$f_0(980)$ and $f_2(1270)$ peaks clearly visible in some ($Q^2$,$x_B$) bins)
sits on top of a $\pi^+\pi^-$ continuum background
(see also fig.~\ref{fig:mmep} where all data have been integrated).

\begin{figure*}
\begin{center}
\includegraphics[width=18cm]{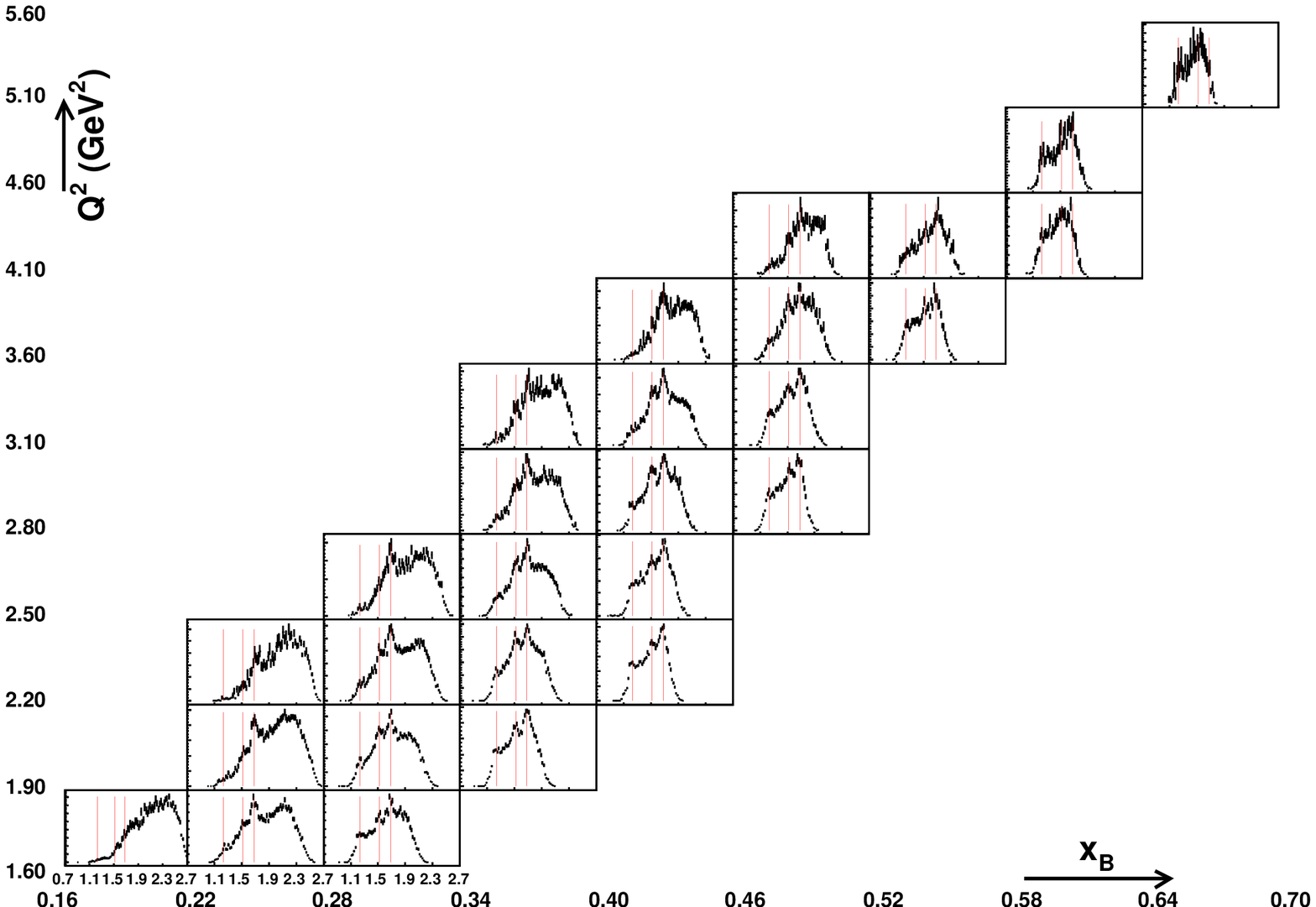}
\caption{Acceptance-corrected $M_{p\pi^-}$ (in GeV) invariant mass distributions for all
our ($Q^2$, $x_B$) bins. The three red lines are located at $M_{X}$ = 1.232, 1.520, 1.680 GeV
corresponding to three well-known resonance regions in the ($p\pi^-$) system.}
\label{fig:IMppim}
\end{center}
\end{figure*}

\begin{figure*}
\begin{center}
\includegraphics[width=18cm]{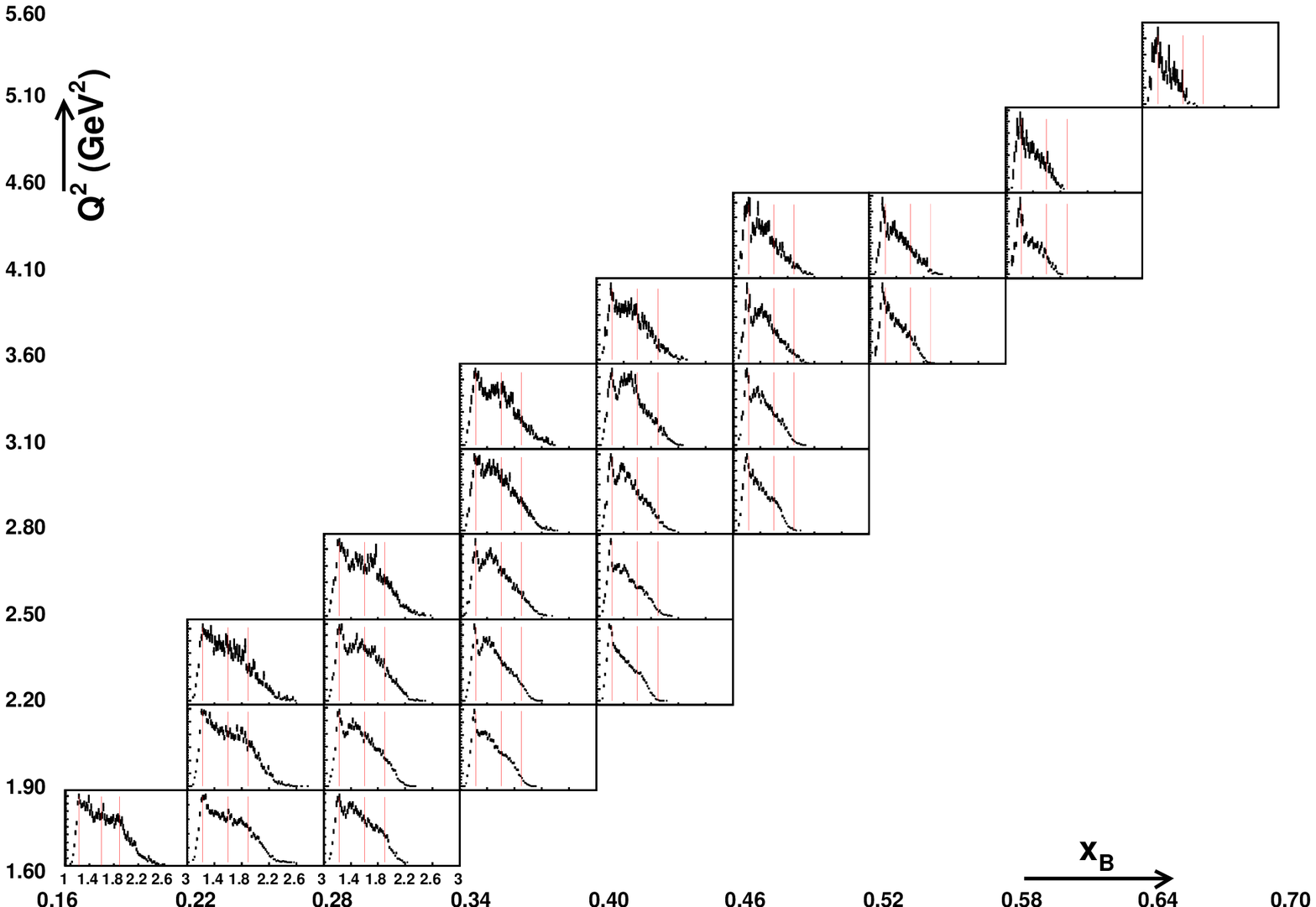}
\caption{Acceptance-corrected $M_{p\pi^+}$ (in GeV) invariant mass distributions for all
our ($Q^2$, $x_B$) bins. The three red lines are located at $M_{X}$ = 
1.232, 1.600, 1.900 GeV
corresponding to the three well-known resonance regions in the ($p\pi^+$) system.}
\label{fig:IMppip}
\end{center}
\end{figure*}

This background can be decomposed, in a first approximation, into the non-resonant 
$ep \rightarrow e^\prime p\pi^+\pi^-$ phase space and the exclusive electroproduction 
of a pion and a nucleon resonance, the latter decaying into a pion and a nucleon, 
such as $ep \rightarrow e^\prime \pi^-\Delta^{++} \hookrightarrow p\pi^+$.
Evidence for this can be seen in figs.~\ref{fig:IMppim} 
and~\ref{fig:IMppip}, which show for all our ($Q^2$, $x_B$) bins 
the acceptance-corrected $p\pi^-$ and $p\pi^+$ invariant-mass
spectra where structures are clearly seen. Most of these
nucleon resonances ($N^*$) are rather well known, the most prominent being the 
$\Delta^{0,++}$(1232), the D$_{13}$(1520) and the F$_{15}$(1680). However,
their production amplitudes with an associated pion  
(i.e. $ep \rightarrow e^\prime \pi N^*\hookrightarrow p\pi$) are mostly unknown.

At low energies ($W<$1.8 GeV) where very few $N^*$
can be produced, a phenomenological model has been 
developed~\cite{mokeev,ripani} based on an effective Lagrangian
where a few $N^*$'s are superposed along with the production
of the $\rho^0$. Such a model could be a strong constraint and guide to 
extract the $\rho^0$ cross section from all the other mechanisms.
However, at the present higher energies, numerous new higher mass $N^*$'s 
appear as shown by the spectra of figs.~\ref{fig:IMppim} and~\ref{fig:IMppip}.
For theoretical calculations, interference effects between all these channels 
are virtually impossible to control and drastically complicate the analysis. 
Therefore this approach cannot be pursued in our case.

At present, it is unrealistic to describe simultaneously 
the $\pi^+\pi^-$, $p\pi^-$ and $p\pi^+$ invariant mass spectra
over our entire phase space because there are too many structures 
varying independently with ($Q^2$, $x_B$) in 
each invariant mass distribution.
 
Therefore, since a complete description of the di-pion mass spectra is 
not available, we have adopted an empirical description of the data using 
non-interfering contributions that, together with a model for the $\rho^0$ 
shape, reproduce the $\pi^+\pi^-$ invariant mass spectrum.

The $\rho^0$ peak is broad and the strength of the non-resonant
$\pi^+\pi^-$ background under it is quite significant, and, 
even more importantly, its nature is unknown. Therefore we must carry out
a non-trivial and model-dependent fitting procedure in order to extract 
the $\gamma^*p \rightarrow p\rho^0$ cross section.

The procedure we have followed consisted of fitting only the
$M_{\pi^+\pi^-}$ distributions for each ($Q^2$, $x_B$) bin in the case of
the integrated cross sections and for each ($Q^2$, $x_B$, $X$) bin, 
where $X$ can be $t$, $\Phi$, $\cos(\theta_{HS})$ 
or $\phi_{HS}$, in the case of the differential cross sections.
It was possible to fit the $M_{\pi^+\pi^-}$ spectra with five contributions:
three Breit-Wigner shapes to describe the three evident mesonic 
$\pi^+\pi^-$ resonant structures of the $\rho^0 (770)$,
$f_0(980)$ and $f_2(1270)$, where the masses in MeV indicated in parentheses
are the values given by the Particle Data Group (PDG)~\cite{PDG}, and 
two smoothed histograms
that are the $M_{\pi^+\pi^-}$ projections of the reactions
$ep \rightarrow e^\prime \pi^-\Delta^{++} \hookrightarrow p\pi^+$ 
and of the non-resonant continuum $ep \rightarrow e^\prime p\pi^+\pi^-$. 
These two latter spectra are calculated by our aforementioned event generator~\cite{genev}.
We now detail these five contributions and explain why they are 
necessary (and sufficient).

We first discuss the contribution of the $\rho^0(770)$ and the way to 
model it. It is well known that simple symmetric Breit-Wigner line shapes 
which are, to first order, used to describe resonances, are too naive to 
reproduce the $\rho$ shape, because of, among other aspects, interference 
effects with the non-resonant $\pi^+\pi^-$ continuum. Several methods
can be found in the literature for treating the $\rho^0$ shape
(see for instance ref.~\cite{bauer} for such a discussion). 
The procedure we adopted was the following:
\begin{itemize}
\item Introduction of an energy-dependent width in order to take into account 
that the $\rho^0$ is an unstable spin-1 particle that decays into two 
spin-0 particles; it is also called a p-wave Breit-Wigner~\cite{Jackson}.
This modified Breit-Wigner reads:
\begin{equation}
BW_{\rho}(M_{\pi^+\pi^-})
=\frac{M_{\rho}\Gamma(M_{\pi^+\pi^-})}{(M_{\rho}^2-M_{\pi^+\pi^-}^2)^2+M_{\rho}^2\Gamma_{\rho}^2(M_{\pi^+\pi^-})}
\label{formbreit1}
\end{equation}
with the energy-dependent width:
\begin{eqnarray}
\Gamma^{\rho}(M_{\pi^+\pi^-})=\Gamma_{\rho}\left(\frac{q}{q_\rho}\right)^{2l+1}\frac{M_{\rho}}{M_{\pi^+\pi^-}},
\end{eqnarray}
where $l=1$ for a p-wave Breit-Wigner, $q$ is the momentum of the decay pion
in the $\rho^0$ center-of-mass frame and $q_\rho$ is equal to $q$ for
$M_{\pi^+\pi^-}=M_{\rho}$:
\begin{eqnarray}
q=\frac{\sqrt{M_{\pi^+\pi^-}^2-4M_{\pi}^2}}{2} ,\,
q_\rho=\frac{\sqrt{M_{\rho}^2-4M_{\pi}^2}}{2}.
\end{eqnarray}

\item Ross and Stodolsky~\cite{RossandStod} and S\"oding~\cite{Soding} 
have shown that the interferences between the
broad $\rho^0$ peak and the important non-resonant $\pi^+\pi^-$
contribution underneath leads to
a skewing of the Breit-Wigner. According to
Ross-Stodolsky, one way to take account of this effect
is to introduce a correction term that consists of multiplying 
the Breit-Wigner formula by an empirical factor that shifts the centroid of 
the $M_{\pi^+\pi^-}$ distribution:
\begin{eqnarray}
BW_{\rho}^{sk.}(M_{\pi^+\pi^-})=BW_{\rho}(M_{\pi^+\pi^-})\left(\frac{M_{\rho}}{M_{\pi^+\pi^-}}\right)^{n_{skew}}.
\label{formbreit2}
\end{eqnarray} 
where $n_{skew}$ is the ``skewing'' parameter. Although 
Ross-Stodolsky have predicted the value of $n_{skew}$ to be 
4, it is often a parameter that is fitted to the data 
since so little is known concerning the interference between the $\rho^0$ 
signal and the $\pi^+\pi^-$ continuum.
\end{itemize}

As is evident in fig.~\ref{fig:MMepXWeighted}, in addition to the 
$\rho^0(770)$, there are two well-known resonant structures in 
the $\pi^+\pi^-$ system: the $f_0(980)$ and 
$f_2(1270)$. Due to the large widths of these mesonic resonances (40 to 100 MeV~\cite{PDG} 
for the $f_0(980)$ and $\approx$ 180 MeV~\cite{PDG} for the $f_2(1270)$), 
it is clearly necessary to include them 
in our $M_{\pi^+\pi^-}$ fit because their contribution can extend into
the $\rho^0$ region, which is itself also broad. 
We also used the formulas of eqs.~\ref{formbreit1}-\ref{formbreit2} for 
these two other mesonic resonant states $f_0(980)$ and $f_2(1270)$
with appropriate parameters $M_{f_0}$, $\Gamma_{f_0}$, $M_{f_2}$, 
$\Gamma_{f_2}$ and took into account their $l=0$ and $l=2$ nature. 

In principle, the only free
parameter to vary in eq.~\ref{formbreit2} should be $n_{skew}$.
However, we have also allowed the central masses and widths of the three mesons
to vary in a very limited range of at most 20 MeV from their nominal values
(see table~\ref{tab:bwparam}).
The motivation for this is that, besides the largely unknown
interference effects between the meson and the $\pi^+\pi^-$ continuum, 
several other effects 
can shift or distort the meson shapes: radiative corrections, binning, 
acceptance corrections, imprecise values for the central 
masses and widths of some of these mesons, etc. 

\begin{table*}	
\begin{center}	
\begin{tabular}{|l|c|c|c|}
\hline	
\hspace*{1.cm} {\bf parameter} & {\bf PDG value} & {\bf min. value} & {\bf max. value} \\
\hline \hline
$\rho^0$ mass $M_{\rho}$ (MeV) & $\approx$ 770 & 750 & 790 \\
$\rho^0$ width $\Gamma_{\rho}$ (MeV) & $\approx$ 150 & 140 & 170 \\
$f_0$ mass $M_{f_0}$ (MeV) & $\approx$ 980 & 970 & 990 \\
$f_0$ width $\Gamma_{f_0}$ (MeV) & 40-100 & 40 & 120 \\
$f_2$ mass $M_{f_2}$ (MeV) & $\approx$ 1275 &  1260 & 1280 \\
$f_2$ width $\Gamma_{f_2}$ (MeV) & $\approx$ 185 & 170 & 200 \\
\hline	
\end{tabular}	
\caption{Range of variations permitted for the parameters to be fitted in
formula~\ref{formbreit2}.} 
\label{tab:bwparam}
\end{center}	
\end{table*}	

Finally, besides the $\rho^0$, $f_0(980)$ and $f_2(1270)$ 
mesons, the two other contributions entering our fit are: 
the $M_{\pi^+\pi^-}$ projections of the reactions 
$ep \rightarrow e^\prime \pi^-\Delta^{++} 
\hookrightarrow p\pi^+$ and the non-resonant continuum (phase space) 
$ep \rightarrow e^\prime p\pi^+\pi^-$. The shapes of these distributions are 
given by our event generator. In particular, the $\Delta^{++}$ 
has the shape of a standard Breit-Wigner in the $p\pi^+$ distribution with
a centroid at 1.232 GeV and a width of 111 MeV~\cite{PDG}. As is obvious
from fig.~\ref{fig:IMppim}, the $\Delta^0$ contribution can be neglected.

In principle, of course, other processes contribute to the $M_{\pi^+\pi^-}$ 
continuum, for instance all of the 
$ep \rightarrow e\pi^-N^* \hookrightarrow p\pi^+$ reactions,
as already mentioned. As a test, we modeled the $p\pi^-$ and $p\pi^+$
invariant mass distributions of figs.~\ref{fig:IMppim} and~\ref{fig:IMppip} 
by adding, at the cross section level, several Breit-Wigners matching the 
structures seen in these figures and identifying them with known $N^*$ 
masses (and widths) that can be found in the PDG. Like for the $\Delta^{++}$,
we introduced their contribution into our fit of the $M_{\pi^+\pi^-}$ distribution using 
our event generator. The conclusion we reached was twofold.
Firstly, this procedure introduced a large number of additional free parameters: 
for each of the extra $N^*$'s, two parameters for the central mass and 
width to vary in the approximate ranges given by the PDG
and one more for the weight/normalization. Secondly, we found that
the $M_{\pi^+\pi^-}$ projected shape of these high mass $N^*$'s 
was very similar to the phase-space 
$M_{\pi^+\pi^-}$ distribution. In other words, in a first approximation,
the phase-space contribution can reflect and absorb the high mass $N^*$'s.
However, the $\Delta^{++}$ contribution to the $M_{\pi^+\pi^-}$
distribution was found to be sufficiently different from the
phase space distribution to be kept as an individual contribution.

\begin{figure*}
\begin{center}
\includegraphics[width=19cm]{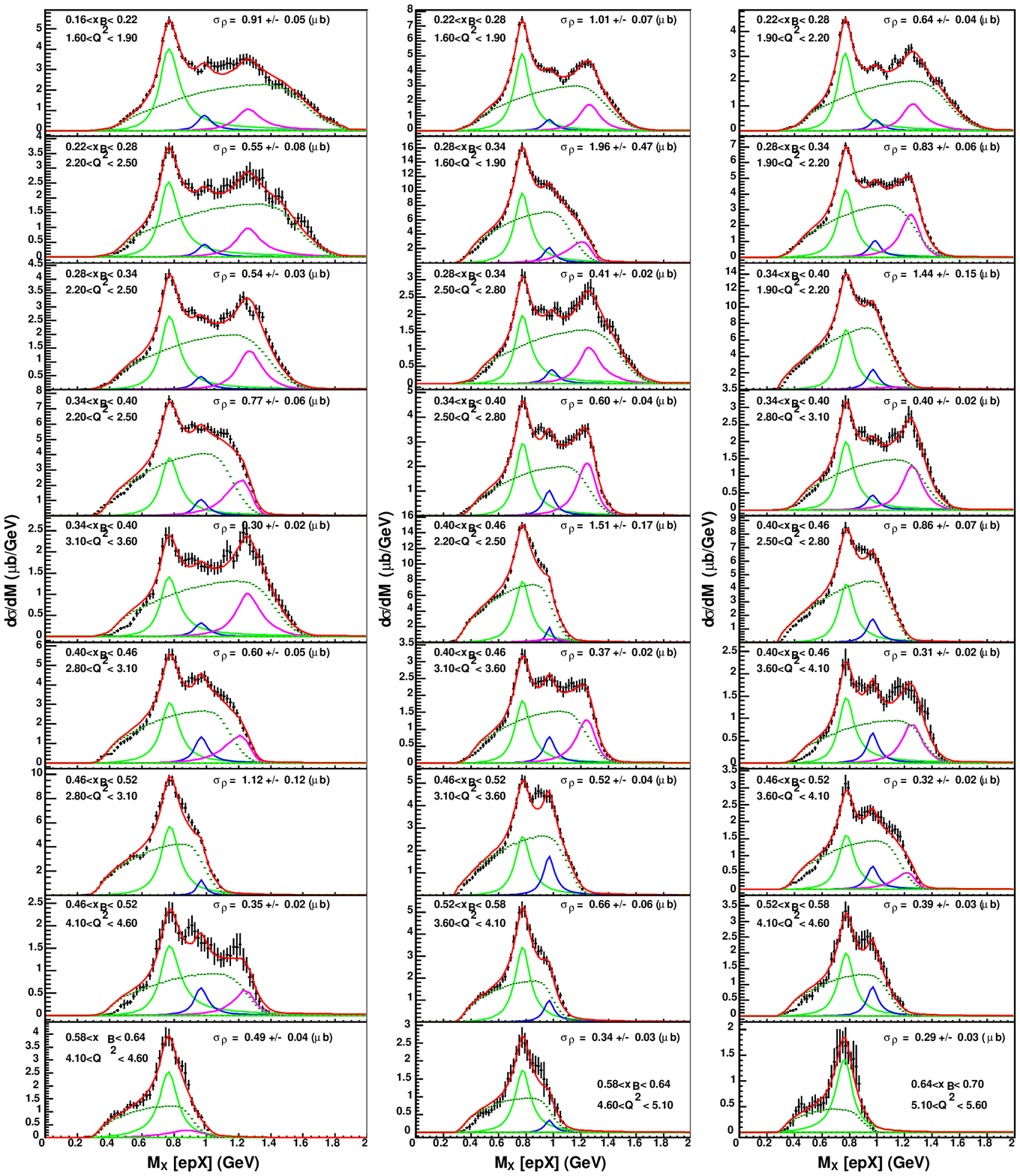}
\caption{Acceptance-corrected $M_{X} [e^\prime pX]$ missing mass distributions , showing our fits. 
	In red: total fit result; in 
	green: $\rho^0$ contribution; in blue: $f_0$ contribution;
	in purple: $f_2$ contribution and in dotted green: $\pi^+\pi^-$
	continuum, which is the sum of the 
	$ep \rightarrow e^\prime \pi^-\Delta^{++} \hookrightarrow p\pi^+$ 
	projections on $M_{\pi^+\pi^-}$ and of the phase space 
	$ep \rightarrow e^\prime p\pi^+\pi^-$ contributions. The error bars on the data points
	are purely statistical. The uncertainties on the cross sections given by the fit
	is also purely statistical. 
	}
\label{fig:sousfondex}
\end{center}
\end{figure*}

To summarize, each ($Q^2$, $x_B$) bin of fig.~\ref{fig:MMepXWeighted}
was fit with the following formula:

\begin{eqnarray}
\frac{dN}{dM_{\pi^+\pi^-}}=&&BW_\rho(M_{\pi^+\pi^-})\nonumber \\
&&+BW_{f_0}(M_{\pi^+\pi^-})+BW_{f_2}(M_{\pi^+\pi^-})\nonumber \\
&&+M_{\Delta^{++}\pi^-}(M_{\pi^+\pi^-})+M_{p\pi^+\pi^-}(M_{\pi^+\pi^-})\nonumber
\\
\label{formbreit3}
\end{eqnarray}

\noindent It involves 14 parameters that are:
\begin{itemize}
\item 1) weight (normalization), 2) central mass, 3) width 
and 4) $n_{skew}$ of $\rho^0$;
\item 5) weight (normalization), 6) central mass, 7) width 
and 8) $n_{skew}$ of $f_0$;
\item 9) weight (normalization), 10) central mass, 11) width 
and 12) $n_{skew}$ of $f_2$;
\item 13) weight (normalization) of $M_{\pi^+\pi^-}$ 
projection of the $ep \rightarrow e^\prime \pi^-\Delta^{++} 
\hookrightarrow p\pi^+$ process; and
\item 14) weight (normalization) of $ep \rightarrow e^\prime p\pi^+\pi^-$
phase space.
\end{itemize}

Fourteen parameters might appear a lot to fit only 
a 1-dimensional distribution. However, on the one hand, six of these (the
central mass and width of the $\rho^0$, $f_0(980)$ and 
$f_2(1270)$ mesons) are quite constrained and are allowed to vary in 
a very limited range. On the other hand this 
simply reflects the complexity and our lack of knowledge of the 
$ep \rightarrow e^\prime p\pi^+\pi^-$ reaction, to which many
unknown, independent though interfering processes contribute; namely:
meson production $ep \rightarrow e^\prime pM^0 
\hookrightarrow \pi^+\pi^-$, $N^*$ production 
$ep \rightarrow e^\prime \pi^-N^{*++} 
\hookrightarrow p\pi^+$, $ep \rightarrow e^\prime \pi^+N^{*0} 
\hookrightarrow p\pi^-$, non-resonant $ep \rightarrow e^\prime p\pi^+\pi^-$,
etc.

Fig.~\ref{fig:sousfondex} shows the result of our fits to the
$M_{\pi^+\pi^-}$ distributions, normalized in terms of the reduced cross sections
of eq.~\ref{eq:sig_ep_eppippim} for all of our ($Q^2$, $x_B$) bins. In a few
cases, the fits do not fully describe the data. For instance, for 
0.46 $< x_B <$ 0.52, 3.10 $< x_B <$ 3.60, the data tend to show a 
``structure" around $M_{\pi^+\pi^-}$=0.9 GeV, i.e. between the
known $\rho^0$ and $f_0$ resonances, which cannot be reproduced by
our fit formula of eq.~\ref{formbreit3}. We attribute this discrepancy
to interference effects not taken into account by our simple fit procedure.
As discussed in more detail in the next subsections, a systematic uncertainty 
of 25\% is assigned to this whole fit procedure which is meant to account, 
among other aspects, for the inadequacies in the model. On all the figures 
that are going to be presented from now on, unless explicitely stated otherwise,
all the error bars associated to our data points will represent the quadratic 
sum of the statistical and systematic errors.

\subsection{Integrated $\rho^0$ cross section}
\label{int}

\begin{figure*}
\begin{center}
\includegraphics[width=15cm]{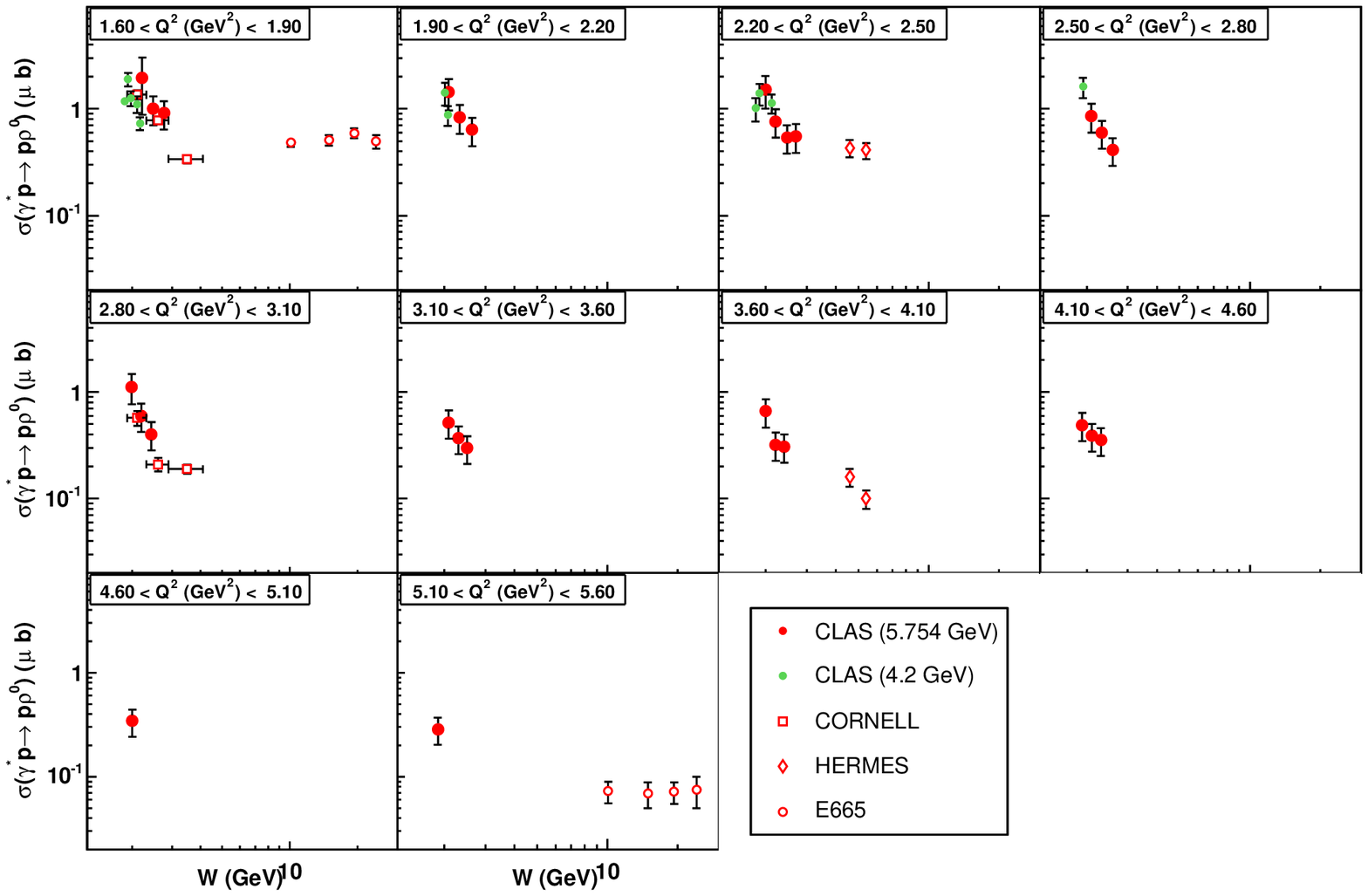}
\caption{Reduced cross sections $\gamma^* p \rightarrow p \rho^0$ 
	as a function of $W$ for constant $Q^2$ bins, 
	in units of $\mu$barn. The error bars of the CLAS data result
	from the quadratic sum of the statistical and systematic uncertainties.	
	The horizontal error bars of the Cornell data indicate
	their $W$ range. The 4.2 GeV CLAS, CORNELL, HERMES and E665 data are 
	respectively from refs.~\cite{cynthia}, \cite{Cassel}, \cite{HERMESrho} 
	and \cite{e665}.}
\label{fig:XsectRhoCompWorldvW}
\end{center}
\end{figure*}

\begin{figure*}
\begin{center}
\includegraphics[width=15cm]{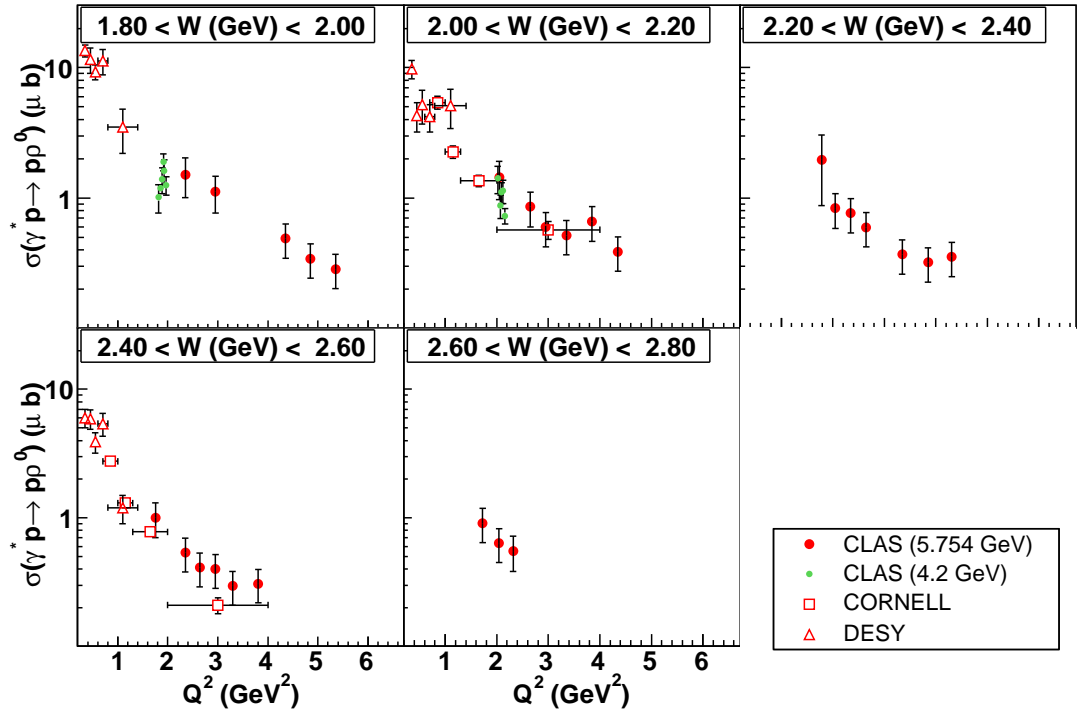}
\caption{Reduced cross sections $\gamma^* p \rightarrow p \rho^0$ 
	as a function of $Q^2$ for constant $W$ bins, 
	in units of $\mu$barn. The 4.2 GeV CLAS, CORNELL and DESY data are 
	respectively from refs.~\cite{cynthia}, \cite{Cassel} 
	and \cite{JoosRho}.
	}
\label{fig:XsectRhoCompWorldvQ2}
\end{center}
\end{figure*}

We use the $\rho^0$ strength (green line) extracted from the distributions 
shown in fig.~\ref{fig:sousfondex} to calculate the cross section. 
Fig.~\ref{fig:XsectRhoCompWorldvW} shows the resulting reduced cross section
$\sigma_{\gamma^*p \rightarrow p\rho^0}$ compared with the world's data  
presented as a function of $W$ for constant $Q^2$ bins.
Fig.~\ref{fig:XsectRhoCompWorldvQ2} shows the reduced cross section
$\sigma_{\gamma^*p \rightarrow p\rho^0}$ compared with the world's data  
presented as a function of $Q^2$ for constant $W$ bins.

With respect to the $\gamma^*p \rightarrow p\pi^+\pi^-$ cross section
that we extracted in the previous section, there is an 
additional source of systematic uncertainty for the 
$\sigma_{\gamma^*p \rightarrow p\rho^0}$ cross section 
that arises from the subtraction procedure described in the previous section.
This contribution is quite difficult to evaluate. It is not so much the quality of
the fit in fig.~\ref{fig:sousfondex} that matters; we have varied
the minimum and maximum limits imposed on the parameters 
in table~\ref{tab:bwparam} and found that the results of
the fits are very stable. The uncertainty arises more from the reliability
and confidence we can assign to the modeling that we have adopted
for the $\rho^0$, $f_0(980)$ 
and $f_2(1270)$ mesons with the skewed Breit-Wigners and for the non-resonant 
continuum $\pi^+\pi^-$ distribution. We have tried several shapes for this
latter continuum. As mentioned in the previous subsection, we introduced
$N^*$ states other than the $\Delta^{++}$. Ultimately, we ended up
finding the fits to be stable at the $\approx$ 20 \% level on average. 
Overall, we cannot take account of any interference effects between the 
$\rho^0$ peak and the non-resonant $\pi^+\pi^-$ continuum. This uncertainty is
of a theoretical nature, and in the absence of sufficient guidance at present, 
we have decided to assign a relatively conservative 25\% systematic uncertainty to 
our extracted $\rho^0$ yields.
We will find some relative justification for this estimation
in the next section when we study the differential distributions, in particular
those of $t$ and $\cos\theta_{HS}$. 

Recently, a partial wave analysis of data of exclusive 
$\pi^+\pi^-$ photoproduction on the proton from CLAS, 
has been carried out~\cite{batta}. This study showed 
that the $\rho^0$ cross sections resulting from this sophisticated method
were consistent with those resulting from simple fits of the 
two-pion invariant mass as we have just described, to a level much lower 
than 25\%. Although no such partial wave analysis has been done
to the present electroproduction data, this photoproduction comparison
gives relative confidence that the 25\% systematic uncertainty that we 
presently assign, is rather conservative.

Coming back to fig.~\ref{fig:XsectRhoCompWorldvW}, we find that our data 
are in general agreement with the other world's data
in regions of overlap. In the upper left plot of 
fig.~\ref{fig:XsectRhoCompWorldvW} (1.60$<Q^2<$1.90 GeV$^2$), 
our CLAS (5.754 GeV) data seem to overestimate the CLAS (4.2 GeV) results, but 
this can certainly be attributed to a kinematic effect due to the 
different beam energies of the two data sets. Indeed, we 
are comparing the total reduced cross sections~: 
$\sigma=\sigma_T+\epsilon\sigma_L$. However, at $W$=2.1 GeV and 
$Q^2$=1.7 GeV$^2$, $\epsilon$=0.53 for a 4.2 GeV beam energy but 
$\epsilon$=0.77 for a 5.754 GeV beam energy. This can readily explain 
the lower CLAS (4.2 GeV) data with respect to the CLAS (5.754 GeV) data.

On this general account, we could have expected that the Cornell data stand
to some extent above the CLAS (5.754 GeV) data since they have been obtained
with an 11.5 GeV beam energy. This is not the case which might indicate 
a slight incompatibility between the Cornell and CLAS 
data. This point, as well as the compatibility of the CLAS (4.2 GeV)  
and CLAS (5.754 GeV) data, will be confirmed in section~\ref{inter} where we
compare the separated longitudinal and transverse cross sections
for which this beam energy kinematical effect is removed. 

\subsection{Differential $\rho^0$ cross sections}
\label{diff}

After having obtained the total $\rho^0$ cross section, we now extract 
the differential cross sections in $t$, $\Phi$, $\cos(\theta_{HS})$
and $\phi_{HS}$. 

Since the data are now binned in an additional variable, 
each bin has fewer statistics, not only for the real data but also 
for the MC data that are necessary to calculate the acceptance 
correction. Bins for which $\eta_w$  is less than 0.6 were rejected,
where, we recall, $\eta_w$ is the correction factor in the acceptance
calculation that was introduced in section~\ref{accep}.
This explains why some holes occur at several instances, in particular
in the $\Phi$ and $\cos(\theta_{HS})$ distributions.

We start by extracting the $d\sigma/dt$ cross section.
Defining $t^{\prime}$ as $t-t_0$, where $t_0$ is the 
maximum $t$ value kinematically allowed for a given ($Q^2$, $x_B$) bin,
we divided the data into 6 bins for $0<-t^{\prime}<1.5\ \hbox{ GeV}^2$ 
and 3 bins for $1.5<-t^{\prime}<3.9\ \hbox{ GeV}^2$.
For each of the ($t$, $Q^2$, $x_B$) bins, we extracted the $\rho^0$ signal
from the $(\pi^+,\pi^-)$ invariant mass spectra using
the fitting procedure previously described.

\begin{figure*}
\begin{center}
\includegraphics[width=18cm]{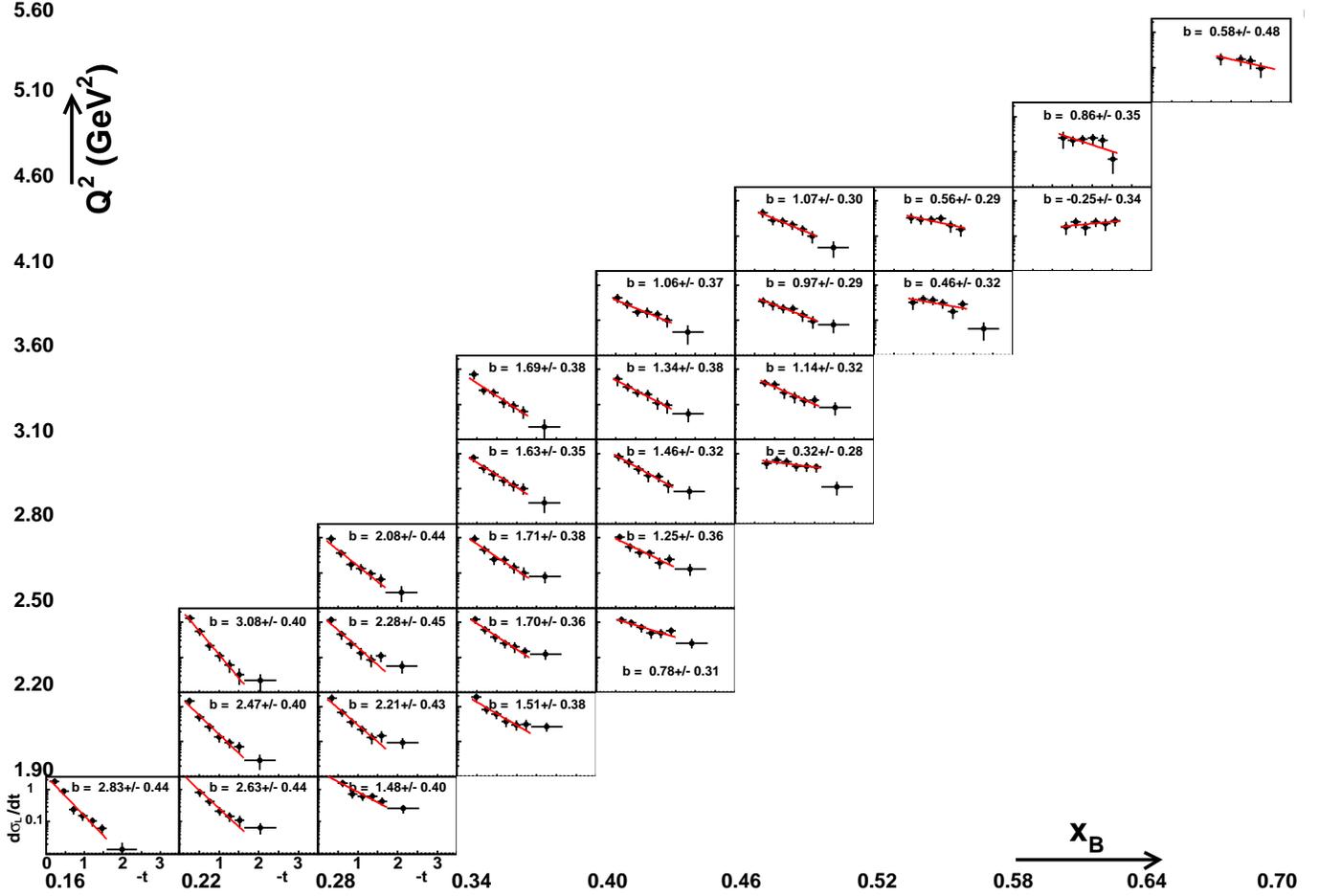}
\caption{Cross section $d\sigma/dt$ (in $\mu$b/GeV$^2$) for all bins in ($Q^2$,$x_B$)
as a function of $-t$ (in GeV$^2$). The red line shows the fit to the function $e^{bt}$
over the limited range $0<-t^{\prime}<1.5\ \hbox{ GeV}^2$.}
\label{fig:exdsigmadtpr}
\end{center}
\end{figure*}

\begin{figure*}
\begin{center}
\includegraphics[width=15cm]{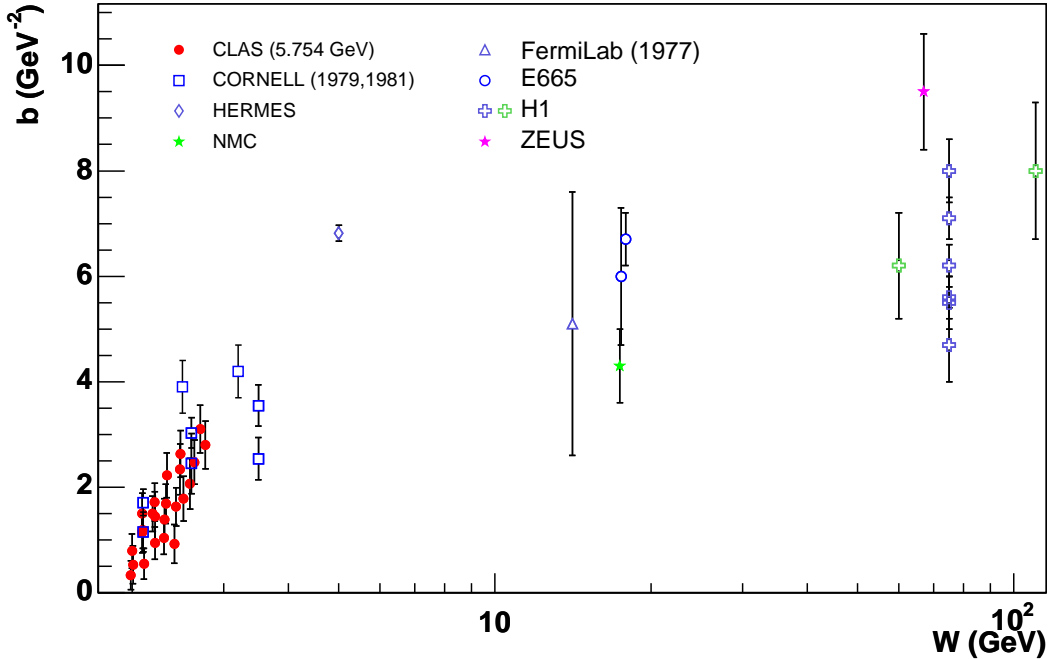}
\caption{Slope $b$ of $d\sigma/dt$ as a function of $W$.
	Data of Cornell~\cite{Cassel,CORNELL},
	HERMES~\cite{HERMESrho}, NMC~\cite{nmc}, Fermilab (1979)~\cite{francis}, 
	E665~\cite{e665}, H1~\cite{h1} and 
	ZEUS~\cite{zeus} are shown for comparison.}
\label{fig:pente_vtpr} 
\end{center}
\end{figure*}

Fig.~\ref{fig:exdsigmadtpr} shows $d\sigma/dt$ for all
our ($Q^2$,$x_B$) bins as a function of $t$.
The general feature of these distributions is that they are of a 
diffractive type, i.e. proportional to $e^{bt}$. 
The values of the slope $b$ are between 0 and 3 GeV$^{-2}$.
They are plotted as a function of $W$ in fig.~\ref{fig:pente_vtpr}
along with the world's data. For the sake of clarity, only the world's data for
$Q^2 >$ 1.5 GeV$^2$ are displayed. For $Q^2 <$ 1.5 GeV$^2$, the data 
show the same trend but with more dispersion. The data exhibit a rise with $W$
until they reach a plateau around $W$= 6 GeV at a $b$ value of 
$\approx$ 7 GeV$^{-2}$. The high-energy experiments 
(H1 and ZEUS) have shown that this saturating value tends to decrease with $Q^2$, 
which is illustrated by the H1 points in fig.~\ref{fig:pente_vtpr}
that correspond to different $Q^2$ values.

By integrating the $d\sigma/dt$ cross section,
we are able to recover at the $\approx$ 20\% level the integrated
cross sections that were presented in section~\ref{int}. 
The agreement is not perfect since for the integrated cross section 
one fits a single full statistics $M_{\pi^+\pi^-}$ spectrum, whereas for 
the differential cross section, one fits several lower statistics $M_{\pi^+\pi^-}$ 
spectra, that are then summed. This relatively good agreement
serves, among other arguments, to justify 
the 25\% systematic uncertainties that we have applied in the non-resonant 
$\pi^+\pi^-$ background subtraction procedure.

We note that the integrated cross sections that 
we have presented so far (and which will be presented in the 
remainder of this article) have been summed over only the domain where we had data
and acceptance. We have not extrapolated our cross
sections beyond the $t$ domain accessed in this experiment, which we deem unsafe
and very model-dependent. Fig.~\ref{fig:exdsigmadtpr} indicates
that this might underestimate some integrated
cross sections for a (very limited) number of ($Q^2$,$x_B$) bins at large $x_B$,
where the $t$ dependence appears rather flat.

\begin{figure*}
\begin{center}
\includegraphics[width=18cm]{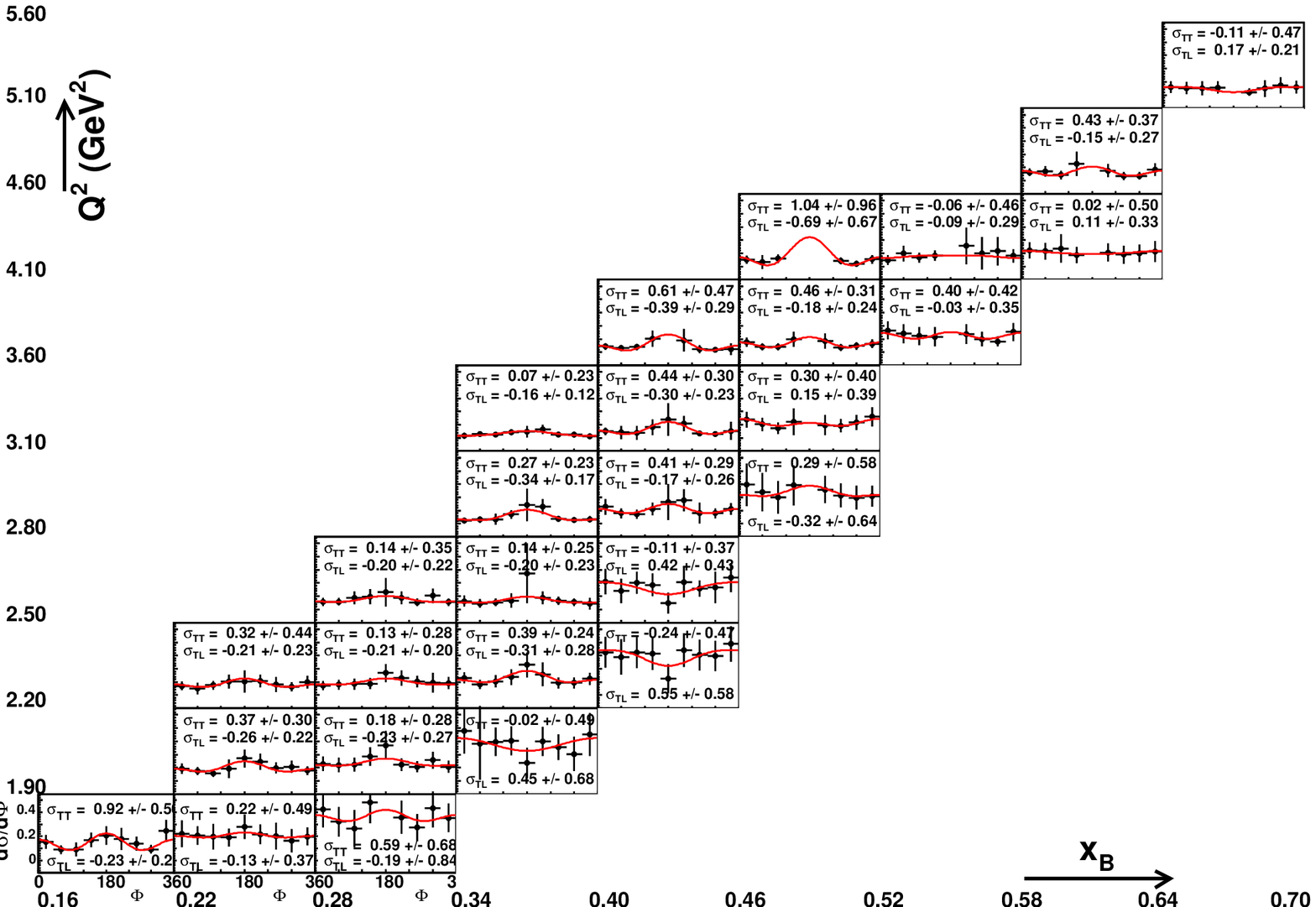}
\caption{Cross section $d\sigma/d\Phi$ (in $\mu$b/rad) for all bins in ($Q^2$,$x_B$) as a function 
of $\Phi$ (in deg.).}
\label{fig:exdsigmadPhi} 
\end{center}
\end{figure*}

We proceeded in the same way to extract $d\sigma/d\Phi$.
All of our ($Q^2$,$x_B$) bins are shown in fig.~\ref{fig:exdsigmadPhi}.
Several of the bins near $\Phi$=180$^o$ are empty or have
large error bars because of very low acceptance in CLAS in this region.   

\begin{figure*}
\begin{center}
\includegraphics[width=15cm]{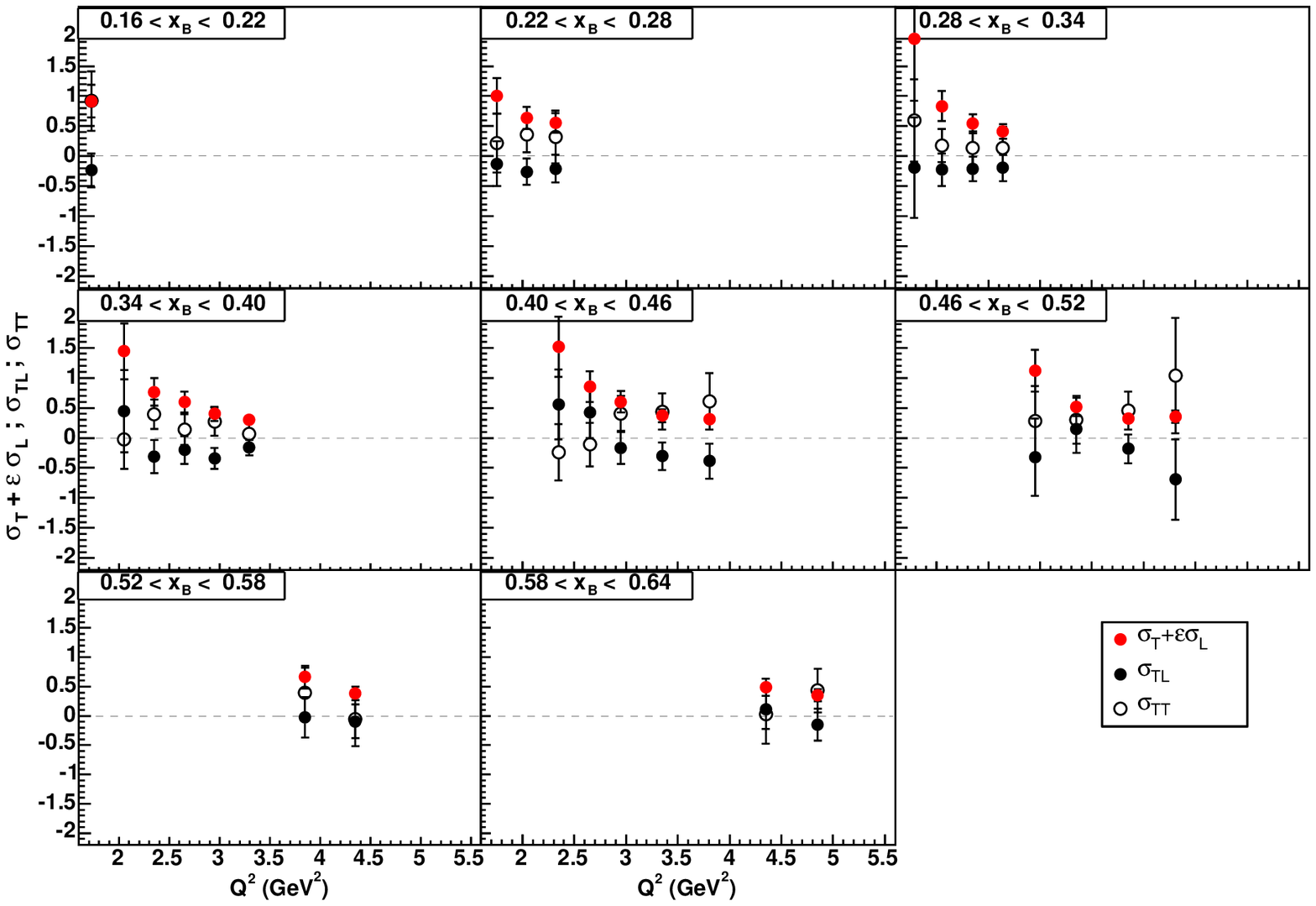}
\caption{Cross sections $\sigma_{T}+\epsilon\sigma_{L}$, $\sigma_{TT}$ and 
$\sigma_{TL}$ (in $\mu$b) 
for the reaction $\gamma^* p \rightarrow p \rho^0$ as a function of $Q^2$ for different 
bins in $x_B$.}
\label{fig:sigTTnsigTL} 
\end{center}
\end{figure*}

These distributions were fitted with the expected $\Phi$ dependence 
for single meson electroproduction:
\begin{eqnarray}
\frac{d\sigma}{d\Phi} = \frac{1}{2\pi} ( && \sigma_T+\epsilon\sigma_L \nonumber \\ 
&&+ \epsilon \cos 2 \Phi \ \sigma_{TT} + 
\sqrt{2\epsilon(1+\epsilon)} \cos \Phi \ \sigma_{TL} )\nonumber \\
\label{eq:sigmaphidep}
\end{eqnarray}
from which we could extract the interference terms $\sigma_{TT}$ and 
$\sigma_{TL}$. The curves in fig.~\ref{fig:exdsigmadPhi} show the 
corresponding fits, and $\sigma_T+\epsilon\sigma_L$, $\sigma_{TT}$
and $\sigma_{TL}$ are displayed in fig.~\ref{fig:sigTTnsigTL}.
If helicity is conserved in the $s$ channel (SCHC), the
interference terms $\sigma_{TT}$ and $\sigma_{TL}$ would vanish.
Most of our extracted values are consistent with 0 within (large) 
error bars, although one clearly cannot make strong claims about 
SCHC at this point.

Turning to the pion decay angles of the $\rho^0$, $\theta_{HS}$ and 
$\phi_{HS}$, they are expected to follow the general and model
independent distribution~\cite{Schillel}:
\begin{eqnarray}
W(\Phi,\cos \theta_{HS},\varphi_{HS}) =\nonumber \\
&& \hspace*{-3cm} \frac{3}{4\pi} \displaystyle\left[ \frac{1}{2}(1-r_{00}^{04}) + \frac{1}{2}(3r_{00}^{04}-1)\cos^2 \theta_{HS} \right] \hspace*{-10cm} \nonumber \\
&& \hspace*{-3cm} -\sqrt{2} \mbox{Re} r_{10}^{04} \sin 2\theta_{HS} \cos \varphi_{HS} - r_{1-1}^{04} \sin^2 \theta_{HS} \cos 2\varphi_{HS}  \nonumber \\
&& \hspace*{-3cm} -\epsilon \cos 2\Phi ( r_{11}^1 \sin^2 \theta_{HS} + r_{00}^1 \cos^2 \theta_{HS} \nonumber \\
&& \hspace*{-3cm} -\sqrt{2} \mbox{Re} r_{10}^1 \sin 2\theta_{HS} \cos \varphi_{HS} - r_{1-1}^1 \sin^2 \theta_{HS} \cos 2\varphi_{HS}) \nonumber \\
&& \hspace*{-3cm} -\epsilon \sin 2\Phi ( \sqrt{2} \mbox{Im} r_{10}^2 \sin 2\theta_{HS} \sin \varphi_{HS} \nonumber \\
&& \hspace*{-3cm} + \mbox{Im} r_{1-1}^2 \sin^2 \theta_{HS} \sin 2\varphi_{HS}) \nonumber \\
&& \hspace*{-3cm} + \sqrt{2\epsilon(1+\epsilon)} \cos \Phi ( r_{11}^5 \sin^2 \theta_{HS} + r_{00}^5 \cos^2 \theta_{HS} \nonumber \\
&& \hspace*{-3cm} - \sqrt{2} \mbox{Re} r_{10}^5 \sin 2\theta_{HS} \cos \varphi_{HS} -r_{1-1}^5 \sin^2 \theta_{HS} \cos 2\varphi_{HS}) \nonumber \\
&& \hspace*{-3cm} + \sqrt{2\epsilon(1+\epsilon)} \sin \Phi ( \sqrt{2} \mbox{Im} r_{10}^6 \sin 2\theta_{HS} \sin \varphi_{HS} \nonumber \\
&& \hspace*{-3cm} + \mbox{Im} r_{1-1}^6 \sin^2 \theta_{HS} \sin 2\varphi_{HS})
\hspace*{-20cm} \left[ \hspace*{+20cm} \displaystyle\right],
\label{eq:omedec}
\end{eqnarray}
where: 
\begin{eqnarray}
r_{ij}^{04} = \frac{\rho_{ij}^0 + \epsilon R \rho_{ij}^4}{1+\epsilon R} \nonumber  \\
&&\hspace*{-3cm}r_{ij}^{\alpha} = \frac{\rho_{ij}^{\alpha}}{1+\epsilon R} \ \ \alpha = 1,2 \nonumber  \\
&&\hspace*{-3cm}r_{ij}^{\alpha} = \sqrt{R} \frac{\rho_{ij}^{\alpha}}{1+\epsilon R} \ \ \alpha = 5,6
\label{eq:rhotor}
\end{eqnarray}
with $R_{\rho}$ equal to the ratio $\sigma_L/\sigma_T$.
 
The parameters $\rho_{ij}^{\alpha}$ are bilinear combinations of 
the helicity amplitudes that describe the $\gamma^*p \to \rho^0 p$ 
transition. They come from a decomposition of the
$3\times 3$ spin density matrix of the $\rho^0$ in a basis of 9
hermitian matrices. The superscript $\alpha$ refers to the virtual photon 
polarization: $\alpha=0$-2 for transverse photons, 
$\alpha=4$ for longitudinal photons, and $\alpha=5$-6 for the interference 
between $L$ and $T$ terms. The subscript refers to the vector meson 
helicity: $i,j=0$ refers to a longitudinal polarization state and
$i,j=-1,1$ to a transverse polarization state. 
For example, $\rho_{00}^0$ is related to the probability of 
the transition between a transverse photon ($\alpha=0$) 
and a longitudinal meson ($i,j=0$) and $\rho_{01}^0$ is an interference term 
between meson helicities 0 and 1 ($i=0, j=1$) produced by a transverse 
photon ($\alpha=0$).

If SCHC applies, then by definition, 
$\rho_{00}^0=0$ and $\rho_{00}^4=1$. Then eq.~\ref{eq:rhotor} leads to a direct 
relation between the measured $r_{00}^{04}$ and the ratio 
$R_{\rho}=\frac{\sigma_L}{\sigma_T}$. In that case,
the longitudinal and transverse cross sections, $\sigma_L$ and $\sigma_T$,
may be extracted from the $\cos\theta_{HS}$ distribution, without 
relying on a delicate Rosenbluth separation.

SCHC can be tested by studying the integrated distributions
$W(\phi_{HS})$ and $W(\Psi)$ (where $\Psi=\phi_{HS}-\Phi$)
over the other decay angles. Integrating 
$W(\Phi, \cos\theta_{HS}, \phi_{HS})$ of eq.~\ref{eq:omedec} 
over $\cos\theta_{HS}$ and $\Phi$ yields:
\begin{equation}
W(\phi_{HS}) = \frac{1}{2\pi} \left[1-2r_{1-1}^{04}\cos 2\phi_{HS} \right],
\label{eq:WvarphiN}
\end{equation}
which isolates $r_{1-1}^{04}$, a density matrix element violating SCHC.
Integrating $W(\Phi, \cos\theta_{HS}, \phi_{HS})$ over 
$\cos\theta_{HS}$ yields:
\begin{equation}
W(\Psi) = \frac{1}{2\pi} \left[1+2\epsilon r^{1}_{1-1}\cos2\Psi \right].
\label{eq:WPsi}
\end{equation}

Another consequence of SCHC is that the $W(\phi_{HS})$ distribution should 
be constant and the $W(\Psi)$ distribution should vary as $\cos2\Psi$ if 
$r^{1}_{1-1}$ is not zero.

\begin{figure*}
\begin{center}
\includegraphics[width=15cm]{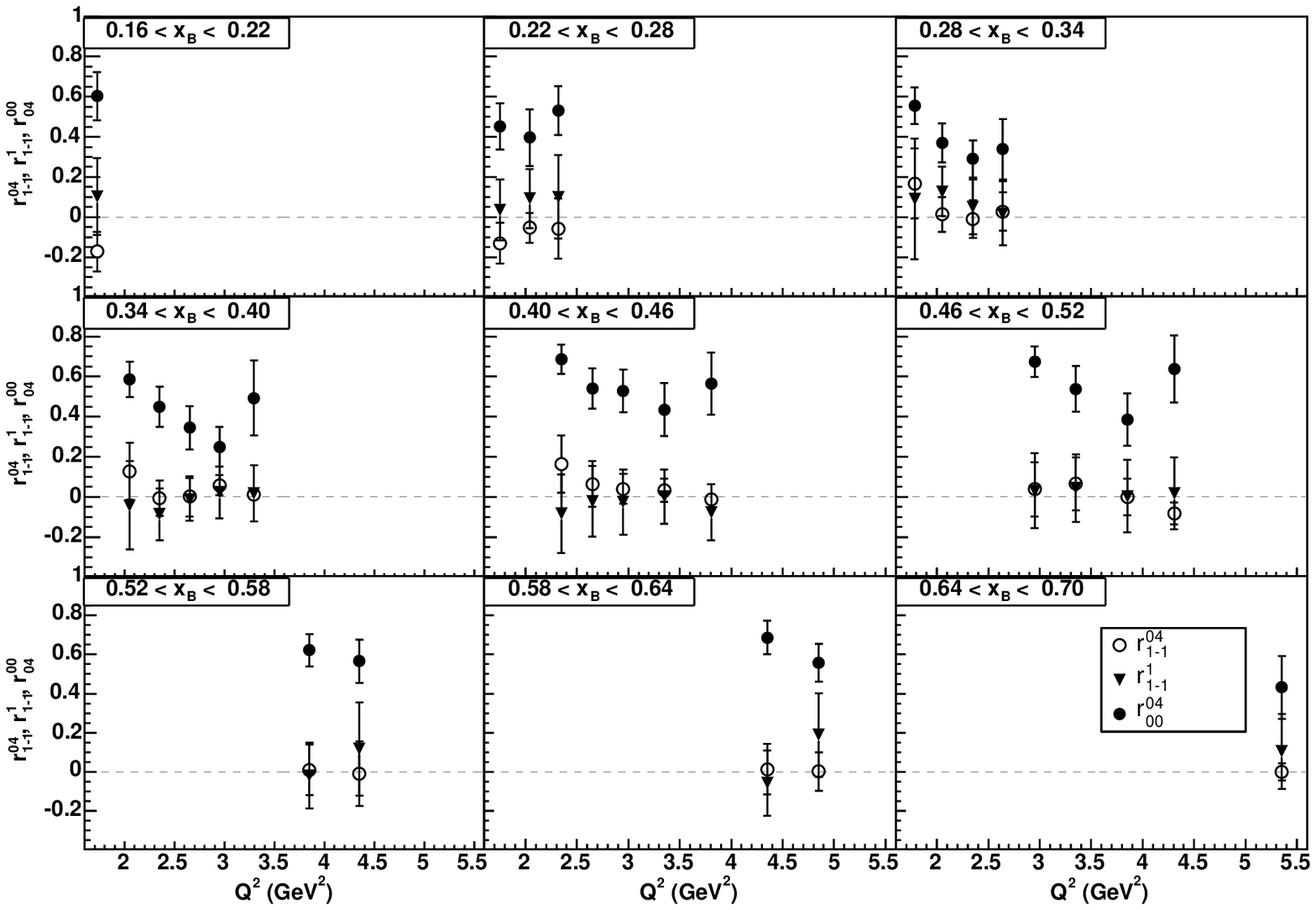}
\caption{Density matrix elements $r^{04}_{1-1}$, $r^{1}_{1-1}$ and $r^{04}_{00}$ for the reaction 
$\gamma^* p \rightarrow p \rho^0$ as a function of $Q^2$ for different 
bins in $x_B$.}
\label{fig:r1-1} 
\end{center}
\end{figure*}

We extracted the $d\sigma/d\phi_{HS}$ and $d\sigma/d\Psi$ cross sections
in the same way as previously mentioned, i.e. by fitting the 
$M_X[e^\prime pX]$ spectra and extracting the $\rho^0$ yield for each 
($Q^2$,$x_B$,$\phi_{HS}$) and ($Q^2$,$x_B$,$\Psi$) bin, respectively.
Fig.~\ref{fig:r1-1} shows the extracted values of $r^{04}_{1-1}$
and $r^{1}_{1-1}$, obtained by fitting $d\sigma/d\phi_{HS}$ and 
$d\sigma/d\Psi$ with the functions of eqs.~\ref{eq:WvarphiN} 
and~\ref{eq:WPsi}, respectively. Basically all the SCHC violating matrix 
elements $r^{04}_{1-1}$ are compatible with 0 (though within large 
uncertainties), which gives some relative confidence in 
the validity of SCHC. In addition, $r^{1}_{1-1}$ is also found to be compatible 
with 0 for all kinematics, although this is not a necessary requirement for SCHC.
This indicates that our $\Psi$ distributions are basically flat.

We finally extract the $r_{00}^{04}$ matrix element from the 
$\cos\theta_{HS}$ distributions, which result from the integration of
$W(\Phi, \cos\theta_{HS}, \phi_{HS})$ of eq.~\ref{eq:omedec} 
over $\phi_{HS}$ and $\Phi$:
\begin{equation}
W(\cos\theta_{HS}) = \frac{3}{8}
\left[(1-r_{00}^{04})+(3r_{00}^{04}-1)\cos^2\theta_{HS} \right].
\label{eq:WcosthetaN}
\end{equation}

\begin{figure}
\begin{center}
\includegraphics[width=7.25cm]{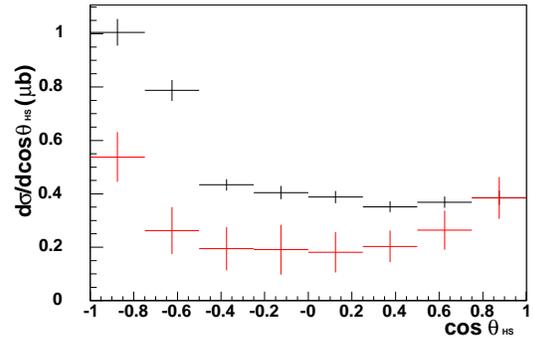}
\caption{An example of a (acceptance corrected) $\cos\theta_{HS}$ 
distribution before (black points) and after (red) the non-resonant 
$\pi^+\pi^-$ background subtraction
($0.58<x_B<0.64$ and $4.10<Q^2<4.60$ GeV$^2$ bin). In this example
the error bars are purely statistical and no systematic uncertainty has been added.
The asymmetry in the red data points between $\cos\theta_{HS}$=-1 and 
$\cos\theta_{HS}$=1 is attributed to some remaining non-resonant 
$\pi^+\pi^-$ background which could not be subtracted by our fitting procedure, 
estimated to lead to a 25\% systematic uncertainty (see section~\ref{back}).}
\label{fig:exsfCOSthetaHS}
\end{center}
\end{figure}

As an example, fig.~\ref{fig:exsfCOSthetaHS} shows a $\cos\theta_{HS}$ 
distribution, before and after the non-resonant 
$\pi^+\pi^-$ background subtraction, for one of our typical
($Q^2$,$x_B$) bins. We note that the unsubtracted distribution is highly 
asymmetrical in $\cos\theta_{HS}$. This is mainly due to the presence of events 
from the $ep \rightarrow e^\prime \Delta^{++}\pi^-$ reaction, whose phase
space is maximum around $\cos\theta_{HS}=1$.

\begin{figure*}
\begin{center}
\includegraphics[width=18cm]{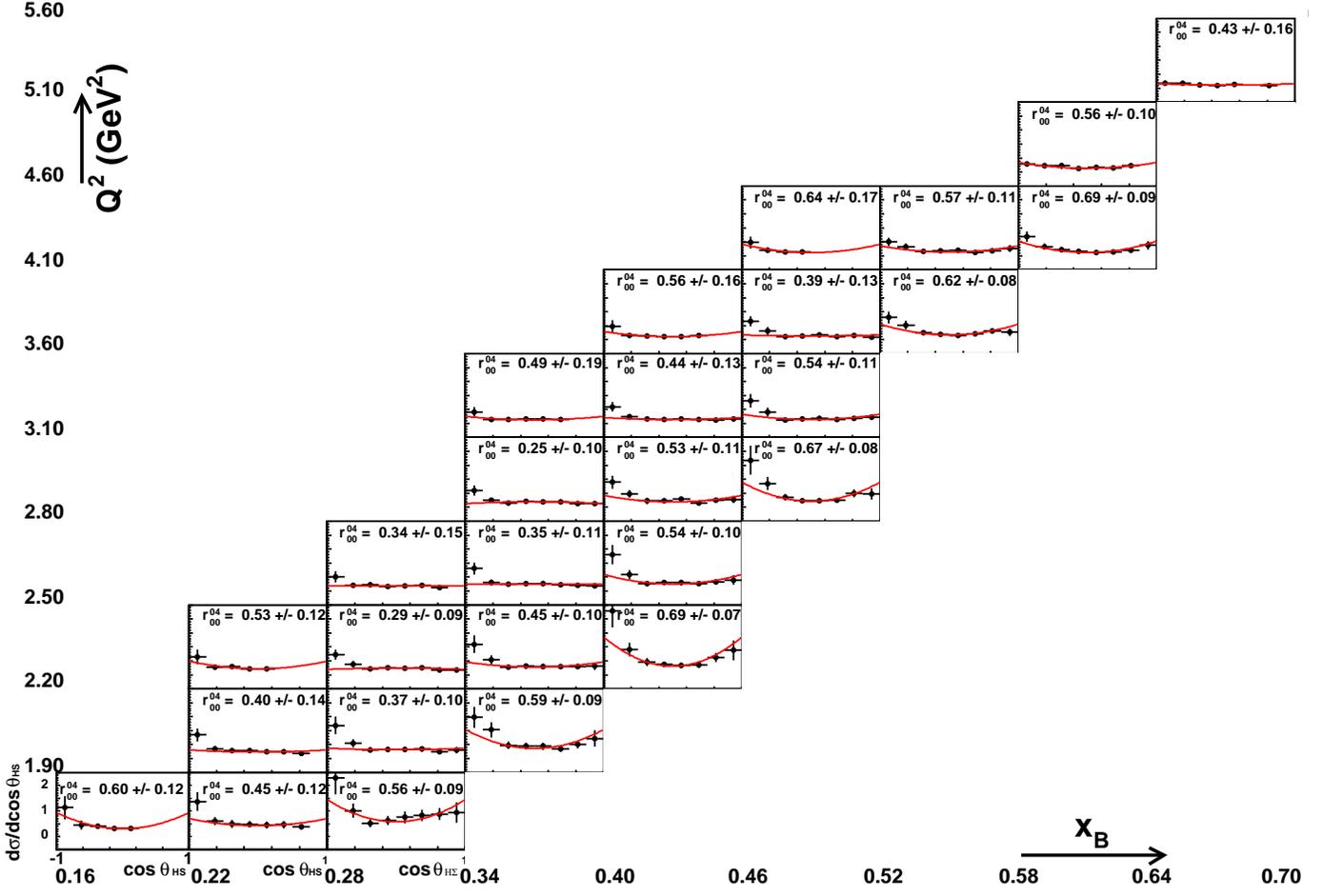}
\caption{Cross section $d\sigma/d\cos(\theta_{HS})$ (in $\mu$b) for all bins in ($Q^2$,$x_B$). 
The red curve corresponds to the fit with function~\ref{eq:WcosthetaN}.
}
\label{fig:exdsigmadcosthHS} 
\end{center}
\end{figure*}

Fig.~\ref{fig:exdsigmadcosthHS} shows the $d\sigma/d\cos(\theta_{HS})$ 
cross sections for all bins in ($Q^2$,$x_B$). Even after the non-resonant 
$\pi^+\pi^-$ background subtraction procedure, some of the aforementioned 
asymmetry in the $\cos(\theta_{HS})$ 
distribution remains at the $\approx$ 25\% level. We attribute this to 
interference effects between the $\rho^0$ channel and its background (mostly, 
$ep \rightarrow e^\prime \pi^-\Delta^{*++} \hookrightarrow p\pi^+$
and non-resonant $ep \rightarrow e^\prime p\pi^+\pi^-$ as already discussed),
which obviously cannot be taken into account when subtracting the 
different channels at the cross-section level as we do.
This $\approx$ 25\% $\cos\theta_{HS}$ asymmetry is a further 
confirmation of the systematic uncertainty associated to the
extraction of the $\rho^0$ signal. Fig.~\ref{fig:r1-1} shows then the resulting 
$r^{04}_{00}$ values.

\subsection{Longitudinal/transverse cross section separation}
\label{landt}

Following the relative verification of the presence of SCHC in the previous
discussion, the ratio $R_{\rho}$ can be determined from: 
\begin{equation}
R_{\rho} = \frac{\sigma_L}{\sigma_T} = \frac{1}{\epsilon}
\frac{r_{00}^{04}}{1-r_{00}^{04}}.
\label{eq:Rrho}
\end{equation}

\begin{figure*}
\begin{center}
\includegraphics[width=15cm]{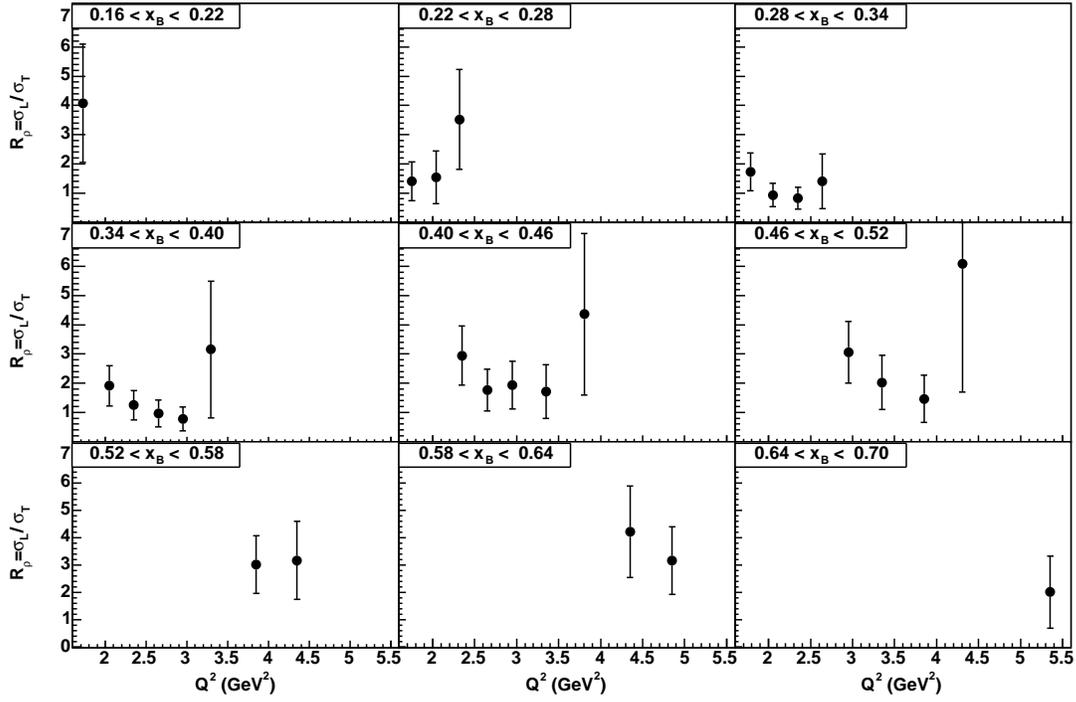}
\caption{The ratio $R_{\rho}$ for the reaction $(\gamma^* p \rightarrow p \rho^0)$ 
as a function of $Q^2$ for different bins in $x_B$.}
\label{fig:Rrho} 
\end{center}
\end{figure*}

Although we cannot claim that our data give strong evidence for SCHC (nor
for its violation), it should be noted that ref.~\cite{hermesLT} mentions that 
eq.~\ref{eq:Rrho} is relatively robust to violations of SCHC.
Fig.~\ref{fig:Rrho} shows $R_\rho$ for all our ($Q^2$,$x_B$) bins,
assuming SCHC, which is an assumption that we 
will keep for the remainder of this analysis.

\begin{figure*}
\begin{center}
\includegraphics[width=15cm]{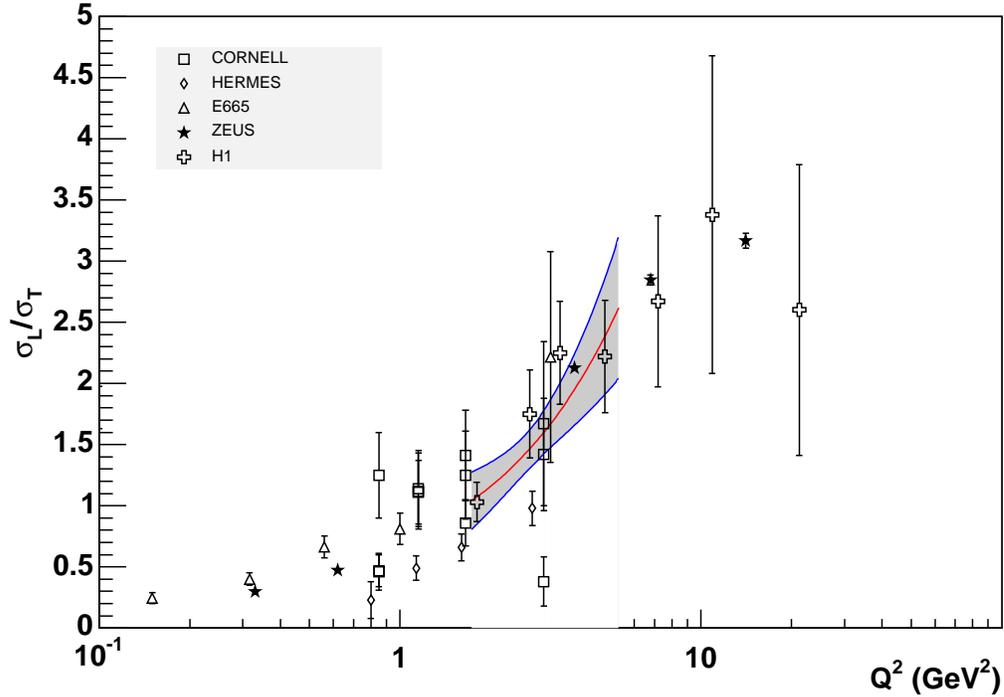}
\caption{World's data for $R_{\rho}$. The grey band represents the fit 
to eq.~\ref{eq:Rrho_fit} using our 27 $R_{\rho}$ points in ($Q^2$,$x_B$). 
The red line shows the central value of the fit 
while the blue lines show the associated uncertainty limits.
The CORNELL, HERMES, E665, H1 and ZEUS data are 
respectively from refs.~\cite{Cassel}, \cite{HERMESrho}, \cite{e665},
\cite{h1} and \cite{zeus}.}
\label{fig:RrhoWD} 
\end{center}
\end{figure*}

We fit our 27 points to a linear function: 

\begin{equation}
R_\rho = a + b Q^2,
\label{eq:Rrho_fit}
\end{equation}

\noindent which yields: $a = 0.281 \pm 0.549$ and $b = 0.439 \pm 0.203$. The
uncertainties on $a$ and $b$ are relatively large but
highly correlated. The normalized correlation coefficent is 0.966.
Fig.~\ref{fig:RrhoWD} shows that the band corresponding to our fit
is in good agreement with the world's data. 

The separated longitudinal and transverse cross sections, 
$\sigma_L$ and $\sigma_T$, are then calculated as:
\begin{equation}
\sigma_T=\frac{\sigma_{\rho}}{(1+\epsilon R_{\rho})}, \,
\sigma_L=\frac{\sigma_{\rho} R_{\rho}}{(1+\epsilon R_{\rho})}.
\end{equation}

We will display these resulting cross sections in section~\ref{theo},
where they will be compared to the theoretical models.

\subsection{Systematic uncertainties}
\label{sys}

There are several sources of systematic uncertainties. The main one stems undeniably
from the fit to extract the $\rho^0$ cross section and subtract the non-resonant 
$\pi^+\pi^-$ continuum, as discussed in section~\ref{back}, and
estimated to be 25\%. We recall that this estimation arises from several
analyses:
\begin{itemize}
\item Fitting the $M_{\pi^+\pi^-}$ distributions of fig.~\ref{fig:sousfondex}
and changing the shapes of the various
inputs and contributions. Other systematic studies on these fits 
were carried out: removing a few data points on the edges of the $M_{\pi^+\pi^-}$ 
spectra to study edge effects, smoothing the histograms to take into
account potential statistical fluctuations, varying the
ranges of the parameters to be fitted (see Table~\ref{tab:bwparam}), etc. 
All in all, we found a stability and robustness of our fits at the $\approx$ 20\% level. 
\item Integrating $d\sigma/dt$ over $t$ (section~\ref{diff})
and comparing it to the total cross sections (section~\ref{int}),
resulting in a $\approx$ 20\% level agreement.
\item Observing an asymmetry at the $\approx$ 25\% level between the
forward and backward angles in the $\cos\theta_{HS}$ distributions
(section~\ref{diff}).
\end{itemize}

A second source of systematic uncertainty stems from the acceptance calculation
which is largely model independent. We have carried out several tests to determine the
associated uncertainty on our procedure. For instance, we have varied the binning of 
the 7-dimensional table (see table~\ref{tab:binning}). As an other test,
we have also varied the input event 
generators: taking for instance only the $ep \rightarrow e^\prime p\pi^+\pi^-$ phase space 
channel or only the $ep \rightarrow e^\prime p\rho^0$ channel. 
Ultimately, we estimated the stability of our acceptance to be at the 15\% level.

We correct the data for radiative effects. These were generated according to 
ref.~\cite{RadCorr}. The approximations used in this calculation may lead to 
systematic uncertainties that we estimate to be of the order of 4\%. 

The determination of the CC efficiencies relies on the assumption that the
distribution of photoelectrons for detected electrons is a generalized Poisson
function and that the shape of this distribution above 4 photoelectrons
is sufficient to determine the whole distribution and to extrapolate to 0.
The maximum error we could make on the integral of the distribution between
0 and 4 photoelectrons is about 25\%. Since the CC inefficiencies are
at most of the order of 6\%, the corresponding systematic uncertainties on the
cross sections is 1.5\%.

The EC efficiency determination relies on the assumption that a particle
with a sufficiently high number of photoelectrons in the CC
was unambiguously an electron. We tried applying several values for this cut to 
estimate the stability of the results and the maximum differences were
found to be on the order of 2\%.

The target length is known to about $\pm$ 1~mm.
The hydrogen density was kept fairly constant through temperature and
pressure stabilization. The determination of the beam integrated charge 
also has a small systematic uncertainty. All this is summarized in 
table~\ref{tab:systerr2}, and leads to a normalization error applicable
to the whole data set. 

\begin{table}[h!]	
\begin{center}	
\begin{tabular}{|l|c|}
\hline	
\hspace*{1.cm} {\bf Source of error} & {\bf Estimated uncertainty} \\
\hline \hline
Fitting procedure & 25\% \\ \hline
CLAS acceptance & 15\% \\ \hline
Radiative corrections & 4\% \\ \hline
CC efficiency & 1.5\% \\ \hline
EC efficiency & 2\% \\ \hline
Target thickness & 2\% \\ \hline
Target density & 1\% \\ \hline
Beam integrated charge & 2\% \\ 
\hline	
\end{tabular}	
\caption{Systematic uncertainties affecting the overall normalization.
The quadratic sum of all theses errors results in a $\approx$ 30\% 
systematic error bar.} 
\label{tab:systerr2}
\end{center}	
\end{table}	

\section{Theoretical interpretation}
\label{theo}

\subsection{The Regge ``hadronic" approach}

The Regge approach consists of understanding exclusive $\rho^0$
electroproduction, above the resonance region and at forward
angles where the cross section is the largest, in terms of exchanges of 
meson ``trajectories" in the $t$-channel. Regge theory generalizes the notion of 
a $t$-channel {\it single} particle exchange to the notion of a {\it family} 
(i.e. trajectory) of particle exchanges. Indeed, mesons, and more broadly hadrons,
appear in general in sequences made of rotational excitations.
Mesons that have the same quantum numbers, except for spin,
seem to align along linear ``trajectories" $\alpha(t)=\alpha(0)+\alpha^\prime t$
that relate their squared mass $-t$ to their spin $\alpha$. This 
leads in the high energy limit to amplitudes proportional to $s^{\alpha(t)}$, 
where $s=W^2$, and therefore total cross sections are
proportional to $s^{\alpha(0)-1}$.

\begin{figure}
\begin{center}
\includegraphics[width=5cm]{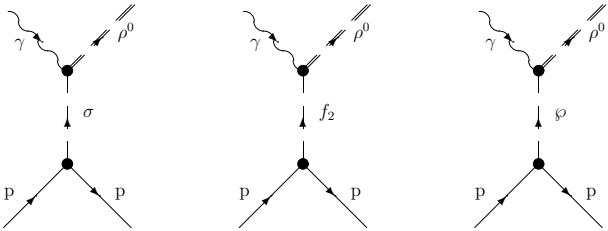}
\caption{Dominant $t$-channel exchange diagrams
 for the reactions $\gamma p \rightarrow p\rho^0$.}
\label{fig:tech} 
\end{center}
\end{figure}

In the following, we will use ``JML" to refer to the latest version~\cite{laget2} 
of the model developed by J.-M. Laget and collaborators~\cite{RgModel2,RgModel3,laget4}. 
The dominant amplitudes correspond to the $t$-channel exchange diagrams 
of fig.~\ref{fig:tech}. Since vector mesons have the same quantum numbers as the 
photon, systems with quantum numbers of the vacuum  can be exchanged. The 
corresponding trajectory is called the pomeron. Although the pomeron contribution 
fully explains $\phi$ meson photo-~\cite{eric} and electroproduction~\cite{joe}, it 
represents only about one third of the cross section in 
the $\gamma^*p\rightarrow p \rho^0$ channel for the energy range covered by our data. 
Here the bulk of the cross section comes from the exchange of the $f_2(1270)$ and 
$\sigma$ mesons. The exchange of the $\pi$ meson, which dominates $\omega$ 
production, contributes very little to the $\rho^0$ production channel.

Amplitudes for the pomeron, $\pi$ and $f_2$ meson exchange diagrams
can be found in ref.~\cite{RgModel2} and for the 
$\sigma$ meson exchange diagram in ref.~\cite{laget4}. In photoproduction, the 
only parameters of the model are the coupling 
constants at the vertices of the diagrams. They are taken from a comprehensive study of 
other independent processes. For instance, the quark-pomeron coupling constant is fixed 
by the analysis of $pp$ scattering at high energy ($W\sim 100$~GeV). 
In electroproduction, a monopole form factor is introduced at 
the $\gamma\pi\rho$ vertex and a dipole from factor is used at the 
$\gamma\sigma\rho$ vertex~\cite{RgModel2}. In this latter reference, 
a dependence on $t$ is given to the cut-off mass that accounts for an increasing 
point-like behavior of the coupling of the photon with the meson when $-t$ 
(and consequently the impact parameter) 
increases. By construction, both the $Q^2$ and the $t$ dependency of the 
$\gamma \wp \rho$ (where $\wp$ stands for the pomeron) 
and the $\gamma f_2 \rho$ vertices are intrinsically part
of the corresponding amplitudes, and no other parameters are included.

With this limited number of parameters, the JML model is able to 
successfully reproduce the main trends of the $W$ and $t$ dependences of the 
total and differential cross sections for the reactions 
$\gamma^* p \rightarrow p\phi$, $\gamma^* p \rightarrow p\omega$ 
and $\gamma^* p \rightarrow p\rho^0$ over the whole $W$ range, i.e. from
threshold up to HERA energies. In order to save computation, in our case, 
the pomeron exchange version of the model has been used instead of the two 
gluon exchange version. In the momentum transfer range of this work, 
i.e. $-t<$ 2 GeV$^2$, the two models lead to almost identical 
results. Overall, as was mentioned before, the pomeron/two-gluon exchange 
contribution does not dominate the cross section in the energy range 
that is accessed in this study ($W$ up to 2.5 GeV). 
 
\subsection{The GPD ``partonic" approach}
\label{gpd_mod}

The JML model was originally built 
for photoproduction, i.e. $Q^2=0$, and was extended to electroproduction 
by introducing form factors to take into account the shorter 
distances probed by the virtual photon, inversely related to $Q^2$. 
We now consider another approach, based on the formalism of 
Generalized Parton Distributions (GPDs), which is valid 
in the so-called Bjorken (or ``Deep Inelastic") regime, i.e.
$Q^2,\nu\to\infty$ with $x_B=\frac{Q^2}{2M\nu}$ finite. An important question is
how low in $Q^2$ this asymptotic formalism can still be applied
or extrapolated?

\begin{figure}[h!]
\begin{center}
\includegraphics[width=5cm]{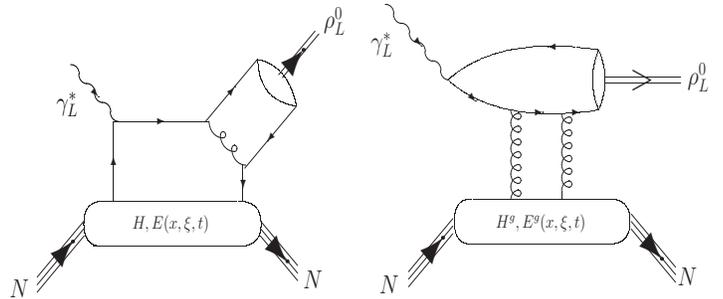}
\caption{The handbag diagrams for (longitudinal) vector meson production.
Quark GPDs are accessed on the left and gluon GPDs are accessed on the right.}
\label{fig:qandg} 
\end{center}
\end{figure}

Collins {\it et al.}~\cite{collins} have shown that the dominant
processes for exclusive meson electroproduction, in the Bjorken regime,
are given by the so-called handbag diagrams represented in
fig.~\ref{fig:qandg}.

The handbag diagrams are based on the notion of factorization
in leading-order pQCD between a hard scattering process, exactly 
calculable in pQCD and a nonperturbative nucleon structure part that is
parametrized by the Generalized Parton Distributions introduced 
by M\"uller {\it et al.}~\cite{muller}, Ji~\cite{ji} and 
Radyushkin~\cite{rady}. For the quark handbag diagram and for vector mesons,
only the two unpolarized GPDs contribute. They are called, using 
Ji's notation, $H$ and $E$, and they depend upon three variables: 
$x$, $\xi$ and $t$. We refer to the rich literature on GPDs 
(see refs.~\cite{goeke,revdiehl,revrady} for recent reviews)
for the full definition of the formalism and of the variables.
For the gluon handbag diagram, the corresponding 
$H^g$ and $E^g$ gluonic GPDs are usually approximated by and reduced to
the forward gluon density $G(x)$.

We recall that for mesons, factorization, 
which is an essential component of the handbag mechanism, is only valid for 
the longitudinal part of the cross section, as the $L$ subscripts on the
photon $\gamma^*_L$ and on the meson $\rho^0$ indicate in fig.~\ref{fig:qandg}. This
is one of the main motivations for separating 
the longitudinal and transverse parts of the cross sections in our data analysis.

We stress that theoretical calculations of exclusive meson production cross sections 
in the QCD factorization and GPD approaches are extremely challenging because 
one needs to address several issues at once: how to model the GPDs; 
how to treat the hard scattering process (the choice of an effective scale 
in $\alpha_s$, the role of QCD corrections, etc.); how to consistently 
combine contributions from meson production in small-size and large-size 
configurations, etc. While in theory these are distinct issues that can be 
discussed separately, in practice they are very much related. Therefore,
the choices and approximations one makes in the treatment of one will 
generally influence the conclusions one draws about the others.

In the following, we will discuss the two particular GK~\cite{gk} and 
VGG~\cite{Vdh1,Vdh2,Vdh3,Vdh4} GPD-based calculations that
provide quantitative results for the {\it longitudinal} 
exclusive $\rho^0$ cross section. Both groups have 
adopted the same approach. They parametrize the $(x,\xi)$ dependence of
the $H$ and $E$ GPDs based on double distributions as proposed in ref.~\cite{RadyDD}, 
(the treatment of the $t$ dependence being different). They
correct the leading order amplitude with
an intrinsic transverse momentum dependence, the so-called $k_\perp$
corrections or, more generally, the modified perturbative approach~\cite{mpa}. 
On this latter point, it is indeed well-known that at high $W$ the 
leading-twist calculations overestimate the data and that the associated 
prediction of $\frac{d\sigma_L}{dt}$ evolving as $\frac{1}{Q^6}$, 
at fixed $x_B$, is not observed in the data 
(see for instance ref.~\cite{koepf}). 

The main difference between the two calculations lies in 
the fact that the GK group has treated the sum of the two handbags 
at the {\it amplitude} level, while the VGG group has treated it 
at the {\it cross section} level, and has therefore neglected
the interference between the two handbag diagrams. We will see this effect in the
next section where we compare our data to the two particular GPD
models we have just introduced.

\subsection{Comparison to data}
\label{inter}

\begin{figure*}  
\begin{center}
\includegraphics[width=15cm]{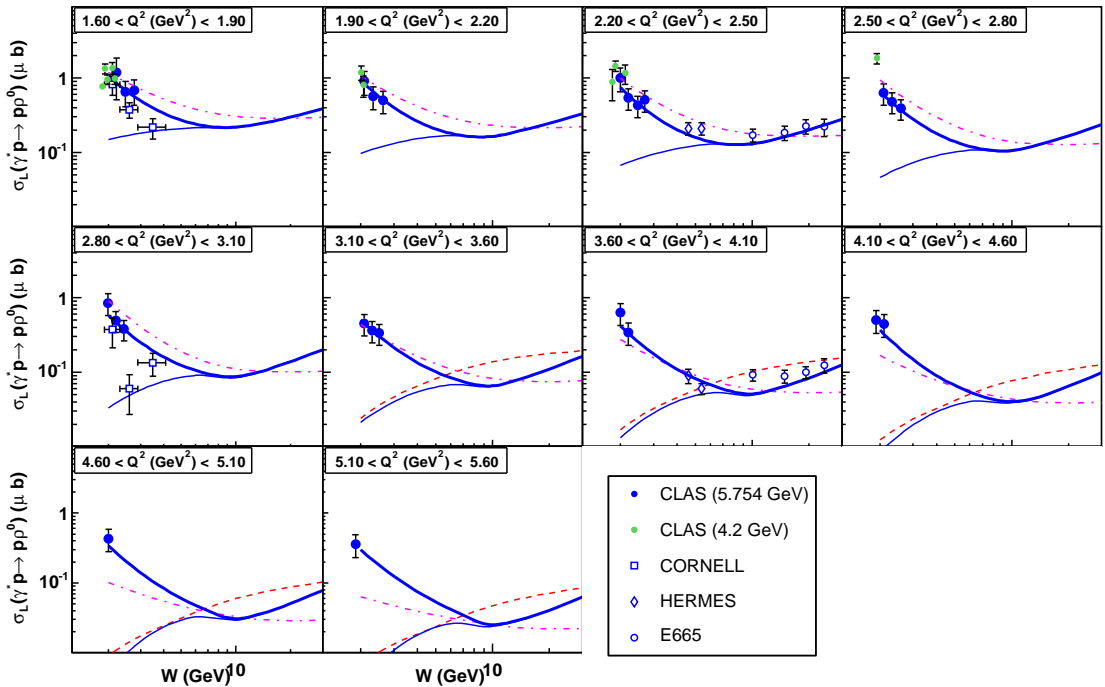}
\caption{World data for the reduced cross sections 
	$\gamma^*_L p \rightarrow p \rho^0_L$ 
	as a function of $W$ for constant $Q^2$ bins, 
	in units of $\mu$barn. The lowest cross section point in the 
	2.80 GeV$^2$ $<Q^2<$ 3.10 GeV$^2$ bin (from CORNELL) corresponds to 
	the low $R_{\rho}$ (=0.38) point in fig.~\ref{fig:RrhoWD}
	and might be unreliable. The dashed curve shows the result of the 
	GK calculation and the thin solid curve shows the result of the VGG calculation.
	Both calculations are based on Double Distributions as proposed in ref.~\cite{RadyDD}
	for the GPD parametrizations and incorporate higher twist effects through
	$k_\perp$ dependence. They differ essentially in summing coherently or not the
	gluon and the quark exchange handbag contributions (see fig.~\ref{fig:qandg}). 
	The thick solid curve is the
	VGG calculation with the addition of the D-term inspired contribution. The dot-dashed
	curve shows the results of the Regge JML calculation.
	The 4.2 GeV CLAS, CORNELL, HERMES and E665 data are 
	respectively from refs.~\cite{cynthia}, \cite{Cassel}, \cite{HERMESrho}
	and \cite{e665}.}
\label{fig:XsectRhoLCompWorldvW}
\end{center}
\end{figure*}

Fig.~\ref{fig:XsectRhoLCompWorldvW} shows our results for
the total {\it longitudinal} cross section for exclusive 
$\rho^0$ electroproduction on the proton as a function of $W$, 
for different $Q^2$ bins, along with the relevant world's data.

The cross sections clearly exhibit two different behaviors as
a function of $W$. At low $W$ the cross sections decrease 
with $W$ and then begin to rise slowly at $W$ $\approx$ 10 GeV. 

In fig.~\ref{fig:XsectRhoLCompWorldvW} the results
of the calculations of the JML, VGG and GK models are also shown. The JML model
(dash-dotted line) reproduces fairly well the two general behaviors just mentioned. 
The drop
of the cross section at low $W$ is due to the $t$-channel 
$\sigma$ and $f_2$ meson exchange diagrams (see fig.~\ref{fig:tech}). 
The intercept $\alpha(0)$ of the $f_2$ trajectory is $\approx 0.5$ 
and therefore the cross sections decrease with energy as 
$\frac{1}{s^{0.5}}$. The flattening of the cross 
section near W $\approx$ 10 GeV comes
from the combined effect of the decreasing $f_2$ contribution and 
the increasing pomeron contribution, whose trajectory has an intercept of 
$\alpha(0)\approx 1+\epsilon$. Although the JML model reproduces 
the general $W$ dependence of the {\it longitudinal} 
exclusive $\rho^0$ cross section relatively well, it drops as a function
of $Q^2$ faster than the data and agrees only up to $Q^2\approx$ 4.10 GeV$^2$.

\begin{figure*}[h!]    
\begin{center}
\includegraphics[width=15cm]{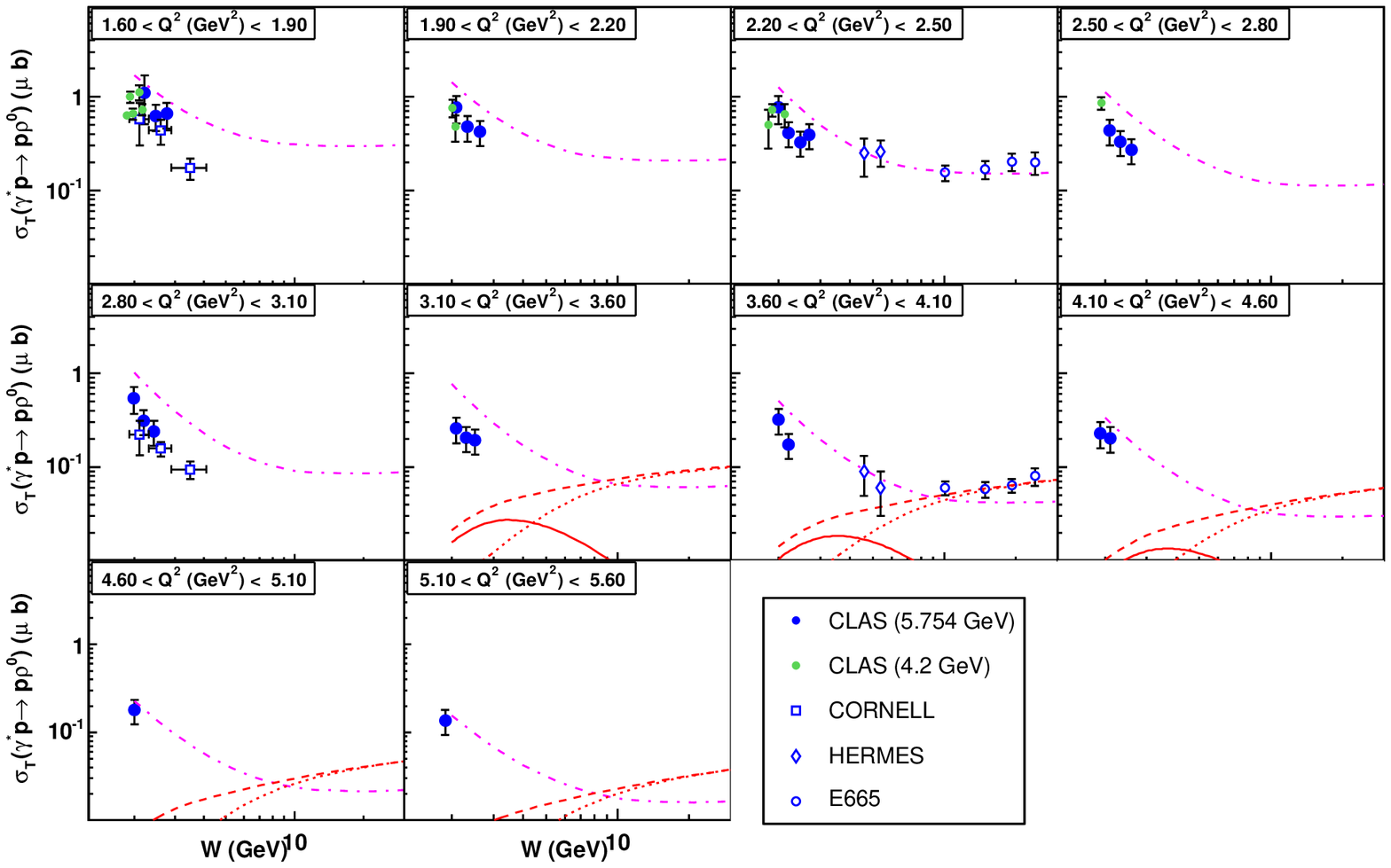}
\caption{World data for the reduced cross sections 
	$\gamma^*_T p \rightarrow p \rho^0_T$ 
	as a function of $W$ for constant $Q^2$ bins, 
	in units of $\mu$barn. The dashed curve shows the result of the 
	GK calculation. The solid curve shows the contribution of the valence
	part of the quark exchange handbag diagram (fig.~\ref{fig:qandg} left) while
	the dotted curve shows the sum of the sea quarks
	part of the quark exchange handbag (fig.~\ref{fig:qandg} left) and of
	the gluon exchange (fig.~\ref{fig:qandg} right) contributions. The dot-dashed
	curve shows the results of the Regge JML calculation. The 4.2 GeV CLAS, 
	CORNELL, HERMES and E665 data are respectively from refs.~\cite{cynthia}, 
	\cite{Cassel}, \cite{HERMESrho} and \cite{e665}.}
\label{fig:XsectRhoTCompWorldvW}
\end{center}
\end{figure*}

We now turn to the GPD approaches. The dashed line shows the result of the 
GK model, while the thin solid line shows the result of the VGG model.
We see that they give a good description 
of the high and intermediate $W$ region, down to $W$ $\approx$ 5 GeV.
This result was already observed by the HERMES 
collaboration~\cite{HERMESrho}. At high $W$ the slow rise of the cross 
section is due to the gluon and sea contributions, while the valence 
quarks contribute only at small $W$ (this decomposition is shown 
in fig.~\ref{fig:XsectRhoTCompWorldvW} when we discuss 
the transverse cross section). 
We see a significant disagreement between the GK and VGG models
at intermediate $W$, which can be clearly explained by the fact 
that, as was mentioned in section~\ref{gpd_mod}, the GK model
takes the interference between the two
handbag diagrams of fig.~\ref{fig:qandg} into account, while the VGG model
sums them {\it incoherently}. This interference is of course maximal
at intermediate $W$'s where the gluon handbag diagram starts to become
significant, while the valence part of the quark handbag diagram
is still significant. The data don't particularly favor
GK over VGG but it is clear that, on purely theoretical grounds,
the GK model is more correct. It is remarkable that, except in this
intermediate $W$ region, i.e. in the high- and low-$W$ regions,
the GK and VGG models are in close agreement. The fact that two independent 
groups with different numerical methods and approximations tend to agree 
gives some relative confidence in the calculations.

At lower $W$ values, where the new CLAS data lie,
it is striking that both the GK and VGG models fail to reproduce 
the data. This discrepancy can reach an order of magnitude at the lowest 
$W$ values. The trend of these particular GPD calculations is to decrease as $W$ 
decreases, whereas the data increase. In the VGG and GK calculations, these trends 
can be understood as follows: GPDs are approximately proportional to 
the forward quark densities $q(x)$. This relation is not so direct
since the quark densities are, in the double distributions ansatz
of ref.~\cite{RadyDD},
convoluted with a meson distribution amplitude but, still, the main trends
remain. Then, as $x$ increases (i.e. $W$ decreases), GPDs tend to go to $0$
since $q(x)\approx (1-x)^3$ for $x$ close to $1$. There might be, according to the
scale, a slight local increase or bump around $x\approx$ 0.3, due to the valence 
contribution, which is indeed clearly apparent in the VGG calculation
shown in fig.~\ref{fig:XsectRhoLCompWorldvW}. However, this variation cannot explain 
an increase of an order of magnitude.

The conclusion on the GPD approach is then two-fold:
\begin{itemize}
\item The handbag is not at all the dominant 
mechanism in the low $W$ valence region and higher twists or so far 
uncontrollable non-perturbative effects obscure the handbag mechanism. 
If so, one has to explain why the (power-corrected) handbag mechanism works in the 
high/intermediate $W$ (i.e. low $x$) domain and, quite abruptly, fails in 
the valence region. Higher twist can certainly depend on energy, but such 
a strong variation with $W$ is certainly puzzling. Also, the explanation 
might simply be of a {\it kinematic} nature. As shown in 
fig.~\ref{fig:exdsigmadtpr}, the minimum value of $\mid t\mid$ increases significantly
with decreasing $W$. For instance, $t_{min}\approx$ 1.6 GeV$^2$ for the 
($0.64<x_B<0.70$, $5.10 <Q^2<5.60$ GeV$^2$) 
bin while $t_{min}\approx$ 0.1 GeV$^2$ for the ($0.16<x_B<0.22$, 
$1.60 <Q^2<1.90$ GeV$^2$) bin. In the handbag formalism, higher twists grow 
with $t$ and this purely kinematic effect provides a natural
source for them. However, more than absolute values, the ratio $\frac{t}{Q^2}$ 
should be relevant, and for the largest $t_{min}$ values, one actually finds
$\frac{t_{min}}{Q^2}=\frac{1.6}{5.35}$, i.e. of the order of 30\%.
More generally, the largest $t_{min}$ values correspond to the largest $Q^2$ values
but, since $Q^2$ increases faster than $t_{min}$ in our kinematics, this actually 
makes the ratio $\frac{t}{Q^2}$ more favorable as $Q^2$ increases.

\item Or the handbag mechanism, which succesfully describes the
region of intermediate and high $W$, is indeed at work in the valence 
region but the way the GPDs are modeled by the VGG and GK groups is incomplete, 
with a significant and fundamental contribution missing, or incorrect.
\end{itemize}

\begin{figure*}
\begin{center}
\includegraphics[width=18cm]{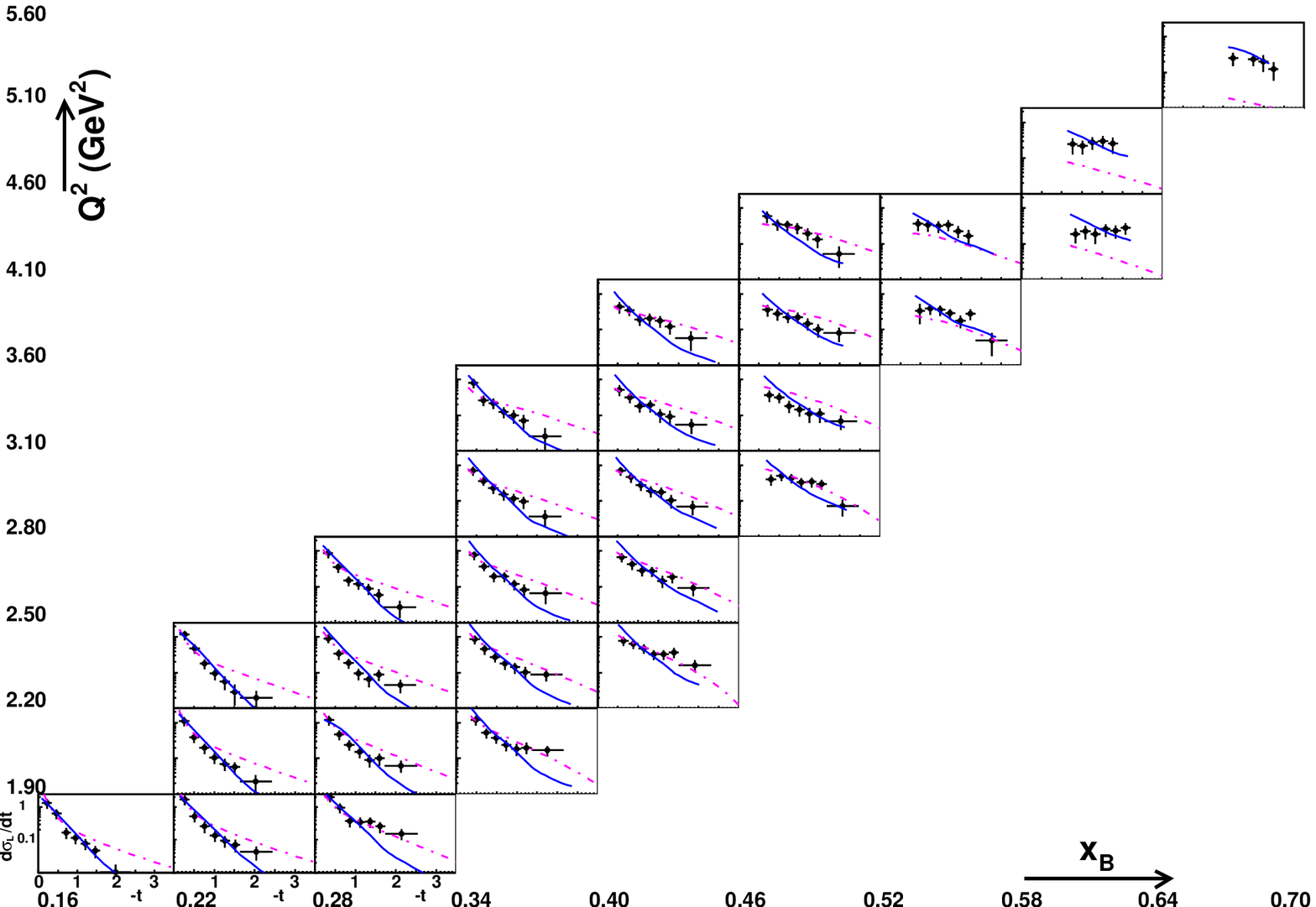}
\caption{Longitudinal cross section $d\sigma_L/dt$ (in $\mu$b/GeV$^2$) for all bins in ($Q^2$,$x_B$)
as a function of $t$ (in GeV$^2$). The thick solid curve represents the result of the VGG calculation
with the addition of the Generalized D-term. The dash-dotted curve is the result of the 
JML model.}
\label{fig:dsdt_theo} 
\end{center}
\end{figure*}

Let us note at this stage that, on general grounds, at large $x_B$, the situation 
is somewhat 
more complex than at small $x_B$. Both real and imaginary parts contribute 
(the so-called ERBL and DGLAP regions), the skewness of the GPDs is substantial 
and non-perturbative effects are expected to play a strong role in determining 
the behavior of the GPDs near $x\to\xi$. Therefore, it is not so clear whether the 
higher twist corrections,
which are already substantial at low $x_B$ through the $k_\perp$ dependence,
have the same character at large $x_B$ and if much can be
concluded about large $x_B$ from the good GPD description of the 
small $x_B$ data.

Therefore, with utmost caution, we quote the suggestion
of ref.~\cite{mick} to add a new (strong) component to the 
standard VGG GPD parametrization, in the form of a $D$-term inspired ansatz,
to reconcile the handbag approach with the data.
We recall that the $D$-term was originally introduced by Polyakov and Weiss~\cite{weiss}
in order to complete the Double Distribution representation of GPDs,
so as to satisfy the polynomiality rule, and that it could be interpreted
as the contribution to the GPDs of the exchange of a $\sigma$ meson in the $t$-channel. 
In ref.~\cite{mick}, the $t$-dependence of the $D$-term was modified (making it, 
effectively,
no longer a $D$-term, properly speaking) and renormalized. One of the motivations for 
this new term was to extend the concept of $q\bar{q}$ components, or $t$-channel meson exchange, 
in GPDs, in a spirit similar to the JML model that explains the strong
rise of the cross section as $W$ decreases by
$t$-channel $\sigma$ and $f_2$ meson exchange processes. The thick solid line of 
fig.~\ref{fig:XsectRhoLCompWorldvW} shows the result of the 
introduction of this new contribution, added coherently to the standard VGG double
distribution parametrization, with its normalization adjusted to the data. 

We insist that this extra contribution is a speculation, which however
does have the merit of providing numerical estimates of the cross section.
Several alternative explanations should also be in order. GPDs can 
obviously be parametrized differently than
in VGG and in GK. It was shown for instance that the spectator model
of Hwang and M\"uller with an overlap representation for the modeling of 
the GPDs~\cite{hwang} produces 
naturally an enhancement at large $x$ compared to the VGG model
that should produce quark exchange cross sections dropping with increasing $W$.
Also, NLO QCD corrections might be more sizeable in this region (see ref.~\cite{kugler}).
Let us finally mention that, in the framework of the VGG model, the Feynman mechanism 
(or overlap diagram, see ref.~\cite{Vdh2} for instance) was calculated but, although it 
has the right $W$ dependence, since it is a real contribution that lives in the 
$\mid x\mid <\xi$ region and therefore dominates at large $x_B$, its numerical contribution, 
which does not rely on any extra parameter, is barely significant. 

In summary, the only conclusion that we allow ourselves to reach at this stage
is that the
two popular GK and VGG models that provide numerical estimations of 
the $ep\to ep\rho^0$ cross sections and which describe well these data at
large $W$ values, fail to describe the present large $x_B$ CLAS data (the normalisation
as well as the $W$ dependence). This fact does not imply however that the handbag
mechanism is not at work in this latter regime as this might simply
be an artifact of the current double distribution parametrization by these
two groups. We reiterate that the necessary consistent 
treatment of GPD modeling, QCD scale setting, and higher-twist effects 
is much more difficult at low $W$ than at high $W$ and renders conclusions
more difficult.

\begin{figure*}
\begin{center}
\includegraphics[width=15cm]{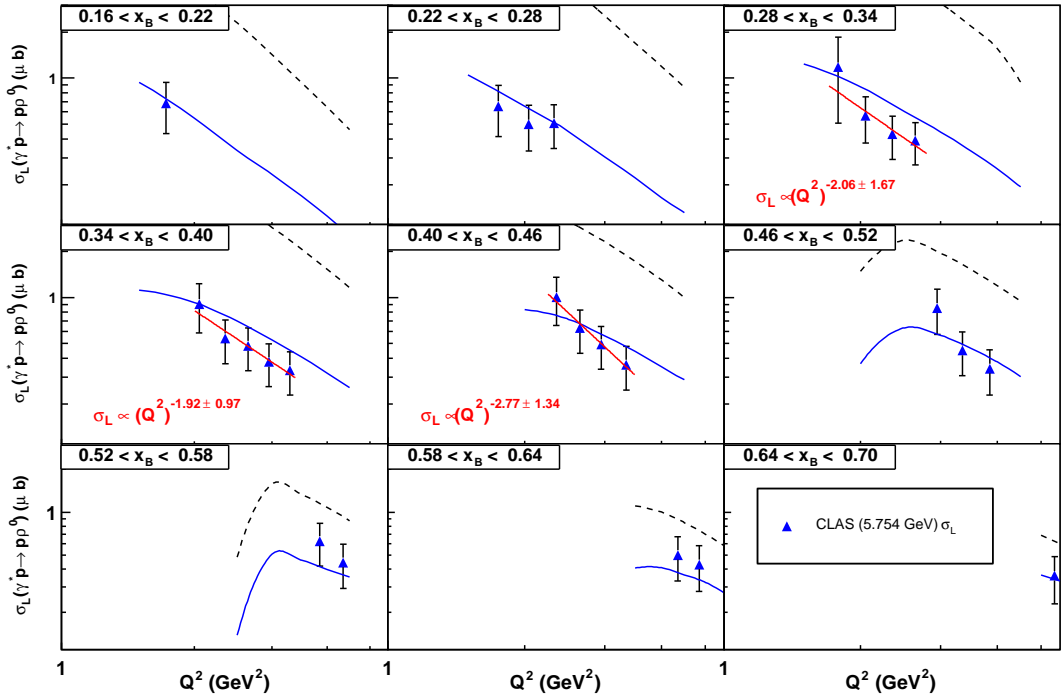}
\caption{Cross sections $\sigma_L$ 
	for the reaction $\gamma^* p \rightarrow p \rho^0$ 
	as a function of $Q^2$ for different bins in $x_B$, 
	in units of $\mu$barn. The solid blue curve is the result of the VGG calculation
	with the Generalized D-term including the $k_\perp$ correction. The dashed curve 
	shows the leading twist (i.e without the $k_\perp$ correction) VGG handbag calculation 
	(with the Generalized D-term). The solid red curve is a fit to the data 
	using a $\frac{1}{Q^{2n}}$ function.}
\label{fig:XsectRhoLandT} 
\end{center}
\end{figure*}

Fig.~\ref{fig:dsdt_theo} shows the {\it longitudinal} differential
cross section $d\sigma_L/dt$ as a function of $t$. The model of ref.~\cite{mick},
inspired partly by Regge theory, naturally explains the decrease of the $t$-slope 
as $x_B$ gets larger. So does the spectator model of 
ref.~\cite{hwang}, which quotes a decrease of the $t$ slope from
$b\approx$ 3.5 GeV$^2$ at $x_B$=0.2 and $Q^2$=2 GeV$^2$ to 
$b\approx$ 1.5 GeV$^2$ at $x_B$=0.6 and $Q^2$=5 GeV$^2$.

Let us mention here that because of the statistics of the data, which in general 
fall quite rapidly with $t$, we have not been able to extract reliably 
the ratio $R_{\rho} = \frac{\sigma_L}{\sigma_T}$ 
as a function of $t$ and that, in a given $(x_B,Q^2)$ bin, the {\it same} $R_{\rho}$
value has been applied to the data over the whole $t$ range. In other words, 
these differential cross sections might need to be corrected 
if the ratio $R_{\rho}$ depends significantly on $t$.

The dash-dotted curves in fig.~\ref{fig:dsdt_theo} show the JML model.
In this approach, the natural $t$ dependence given by the Regge formula $s^{\alpha (t)}$
is too sharp, as for the $H$ and $E$ GPD case. Thus, $t$-dependent form 
factors have to be introduced at the electromagnetic vertices of the diagrams of 
fig.~\ref{fig:tech}, according to the procedure and motivations of ref.~\cite{laget2}.

Fig.~\ref{fig:XsectRhoLandT} shows the $Q^2$ dependence of the 
longitudinal $\rho^0$ cross section $\sigma_L$ for different $x_B$ values.
When more than three points are present in an $x_B$ bin, we fit
the $Q^2$ dependence to the function $\frac{1}{Q^{2n}}$, with the extracted values $n$ 
displayed in the figure. We recall that the handbag formalism predicts a value of 3
for $n$ at asymptotically large $Q^2$ values. However, a smaller coefficient, 
i.e. a flatter $Q^2$ dependence, is expected at these low $Q^2$ values due to 
preasymptotic ($k_\perp$) effects.
The thick solid (blue) curves in fig.~\ref{fig:XsectRhoLandT} show the results of the VGG 
calculation including the extra aforementioned term.
The magnitude and shape of the data are reasonably reproduced. The $k_\perp$ effects
in the calculation flatten the $Q^2$ slope of the cross section. For comparison, 
the asymptotic result, i.e. without $k_\perp$ effects, is shown as the dashed (black)
curve in fig.~\ref{fig:XsectRhoLandT}. One sees that its normalization is of course higher.
Indeed, $k_\perp$ effects reduce the cross section by a factor 2 to 5 depending
on $x_B$ and $Q^2$. Also, the $Q^2$ dependence of the asymptotic
result is steeper, precisely $\frac{1}{Q^6}$. At asymptotically large $Q^2$ values, the two
calculations, i.e. with and without $k_\perp$ effects, are expected to agree.

To complete the interpretation of our data, we finally 
turn to the transverse part of the cross section.
We show in fig.~\ref{fig:XsectRhoTCompWorldvW} the $W$ dependence
of the $\gamma^*_T p \rightarrow p \rho^0_T$ cross section for different $Q^2$ bins.
The JML model (dot-dashed curve) once again reproduces the general shape 
of the $W$ dependence of this cross section. Though, quantitatively, it seems to 
overestimate by $\approx$ 30\% the CLAS data at low $Q^2$, where it is expected 
to be the most valid. The agreement for the longitudinal 
part of the cross section was much better for this kinematical region.

\begin{figure*}
\begin{center}
\includegraphics[width=18cm]{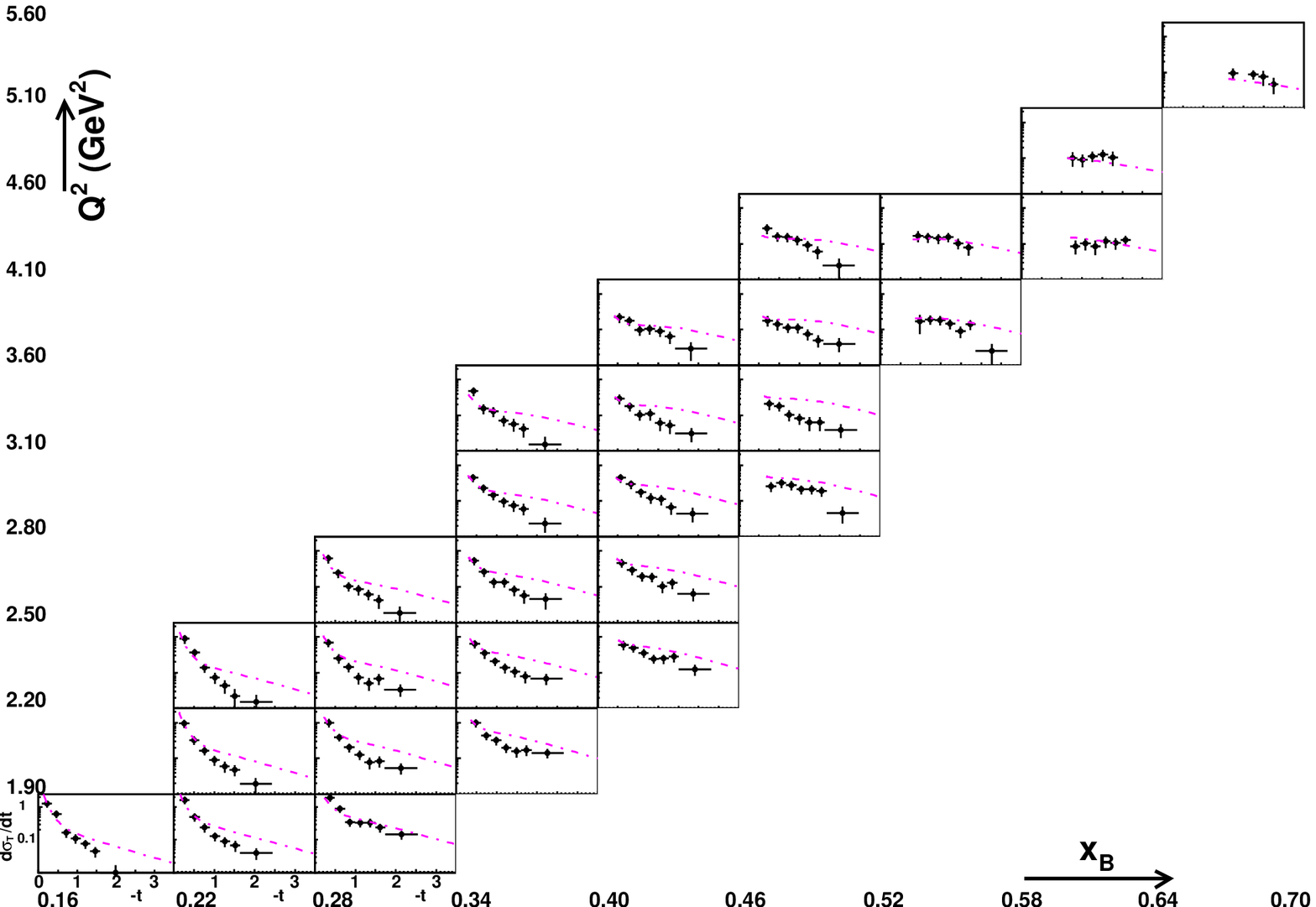}
\caption{Cross section $d\sigma_T/dt$ (in $\mu$b/GeV$^2$) for all bins in ($Q^2$,$x_B$)
as a function of $t$ (in GeV$^2$). The dash-dotted curve is the result of the 
JML model.}
\label{fig:dstdt_theo} 
\end{center}
\end{figure*}

The transverse part of the cross section doesn't lend itself straighforwardly to
a GPD interpretation since it is higher twist. However, the GK group,
taking into account $k_\perp$ effects, has been able to extend its analysis of the 
longitudinal cross section to the transverse case~\cite{gk2}. They showed
that retaining the quark transverse momenta regularizes the infrared singularities
occuring in the transverse process. The dashed curve in fig.~\ref{fig:XsectRhoTCompWorldvW} 
shows the result of this calculation. The high energy E665 data, and to some extent,
the HERMES data are well reproduced, thus comfirming the approach. However, as for the 
longitudinal cross section, the low $W$ CLAS data are completely underestimated. The 
VGG model has not yet 
been extended to the transverse case but it is clearly expected that the addition
of the Generalized D-term to the transverse process produces the same effect
as for the longitudinal one and might explain the rise of the cross section at low $W$.

Finally, fig.~\ref{fig:dstdt_theo} shows the transverse differential cross section
$d\sigma_T/dt$ compared to the JML model. One finds that the JML model tends to 
overestimate the experimental cross sections
at the $\approx$ 30\% level, especially at large $t$ values.

\section{Summary}
\label{conc}

Using the CLAS detector at JLab, we have collected 
the largest ever set of data for the $e p\to e^\prime p \rho^0$
reaction in the valence region. We have presented the $Q^2$ and $x_B$ (and $W$)
dependences of the total, longitudinal and transverse cross sections, as well as
of the differential cross section in $t$.

The unique features that we have observed are:
\begin{itemize}
\item The $W$ dependence of our data shows a clear decrease of the cross sections
with increasing $W$, in contrast to the higher $W$ data (HERMES, H1, ZEUS, E665),
which show cross sections that tend to be flat or slowly rising with $W$,
\item The $t$ dependence exhibits a varying slope with energy: the slope
increases as $x_B$ decreases. In particular, our $t$ dependences are almost flat 
at our largest $x_B$ values,
\item The cross sections decrease with $Q^2$
as approximatively $\frac{1}{Q^4}$, i.e. in a flatter way than what is predicted
by the asymptotic handbag diagram.
\end{itemize}

These data and features can be interpreted in two ways:
\begin{itemize}
\item In terms of hadronic degrees of freedom, i.e. meson trajectory exchanges
in the $t$-channel, following the JML Regge model. In order to reproduce the rather 
flat $t$ dependences varying with $Q^2$ and $x_B$, electromagnetic form factors varying
with $Q^2$ and $t$ are necessary. For the longitudinal part of the cross section,
good agreement with the data is found up to $Q^2\approx$ 4.10 GeV$^2$. For the
transverse part of the cross section, there seems to be an overestimation 
(by $\approx$ 30\%) of the cross section.
\item In terms of partonic degrees of freedom, i.e. quark handbag diagrams and
Generalized Parton Distributions. However, the GK and VGG calculations cannot 
provide the right $W$ dependence of the cross section. This does not
imply that GPDs cannot be accessed through exclusive $\rho^0$ electroproduction 
in the valence region but that possibly the way double distributions 
are modeled or the hard scattering amplitude is calculated in these 
two particular approaches should be modified or revisited.
We stress that in exclusive meson electroproduction the GPD modeling problem 
is convoluted with other issues such as the treatment of the QCD scale setting, 
higher-twist effects, the meson distribution amplitude, etc. ,
rendering conclusions difficult. The present data 
will provide important input to improve our understanding of these fundamental 
QCD issues.
\end{itemize}

\vskip1.cm

\textbf{Acknowledgements}

\vskip1.cm

We would like to thank the staff of the Accelerator and Physics
Divisions at Jefferson Lab who made this experiment possible.
It is also a pleasure to thank S. Goloskokov, P. Kroll, D. M\"uller,
M. Vanderhaeghen and C. Weiss for insightful discussions.
Acknowledgments for the support of this experiment go also to the Italian
Istituto Nazionale di Fisica Nucleare, the French Centre National de la
Recherche Scientifique and Commissariat \`a l'Energie Atomique, the UK
Engineering and Physical Science Research Council, the
U.S. Department of Energy and the National Science Foundation, and the
Korea Research Foundation. The Southeastern Universities
Research Association (SURA) operated the Thomas Jefferson National
Accelerator Facility under U.S. Department of Energy contract
DE-AC05-84ER40150.


\end{document}